\documentclass[conference]{IEEEtran}
\IEEEoverridecommandlockouts
\usepackage{cite}
\usepackage{hyperref}
\usepackage{amsmath,amssymb,amsfonts}
\usepackage{algorithmic}
\usepackage{graphicx}
\usepackage{textcomp}
\usepackage{xcolor}

\usepackage[utf8]{inputenc}
\usepackage[T1]{fontenc}

\usepackage{tikz}
\usepackage{amsmath}
\usepackage{algorithmic}
\usepackage{algorithm}
\usepackage{booktabs}
\usepackage{array}
\usepackage{graphicx}
\usepackage{booktabs}
\usepackage{multirow}
\usepackage{balance}
\usepackage[inkscapelatex=false]{svg}

\def\BibTeX{{\rm B\kern-.05em{\sc i\kern-.025em b}\kern-.08em
    T\kern-.1667em\lower.7ex\hbox{E}\kern-.125emX}}

\begin{document}

\title{ZeroLog: Zero-Label Generalizable Cross-System Log-based Anomaly Detection
\thanks{* Corresponding Authors}
}

\author{\IEEEauthorblockN{Xinlong Zhao}
\IEEEauthorblockA{\textit{School of Software and Microelectronics} \\
\textit{Peking University}\\
Beijing, China \\
xlzhao25@stu.pku.edu.cn}
\and
\IEEEauthorblockN{Tong Jia *}
\IEEEauthorblockA{\textit{Institute for Artificial Intelligence} \\
\textit{Peking University}\\
Beijing, China \\
jia.tong@pku.edu.cn}
\and
\IEEEauthorblockN{Minghua He}
\IEEEauthorblockA{\textit{School of Software and Microelectronics} \\
\textit{Peking University}\\
Beijing, China \\
hemh2120@stu.pku.edu.cn}
\and
\IEEEauthorblockN{Ying Li *}
\IEEEauthorblockA{\textit{National Engineering Research Center for Software Engineering} \\
\textit{Peking University}\\
Beijing, China \\
li.ying@pku.edu.cn}
\and
\IEEEauthorblockN{Gang Huang}
\IEEEauthorblockA{\textit{Institute for Artificial Intelligence} \\
\textit{Peking University}\\
Beijing, China \\
hg@pku.edu.cn}

}

\maketitle

\begin{abstract}
Log-based anomaly detection is an important task in ensuring the stability and reliability of software systems. One of the key problems in this task is the lack of labeled logs. Existing works usually leverage large-scale labeled logs from mature systems to train an anomaly detection model of a target system based on the idea of transfer learning. However, these works still require a certain number of labeled logs from the target system. In this paper, we take a step forward and study a valuable yet underexplored setting: zero-label cross-system log-based anomaly detection, that is, no labeled logs are available in the target system. Specifically, we propose ZeroLog, a system-agnostic representation meta-learning method that enables cross-system log-based anomaly detection under zero-label conditions. To achieve this, we leverage unsupervised domain adaptation to perform adversarial training between the source and target domains, aiming to learn system-agnostic general feature representations. By employing meta-learning, the learned representations are further generalized to the target system without any target labels. Experimental results on three public log datasets from different systems show that ZeroLog reaches over 80\% F1-score without labels, comparable to state-of-the-art cross-system methods trained with labeled logs, and outperforms existing methods under zero-label conditions.

\end{abstract}


\section{Introduction}
As the scale and complexity of software systems continue to grow, the frequency of failures has shown an upward trend. Ensuring the reliability and stability of systems has become one of the core challenges for their successful operation. System logs, which record key events and state changes during system operation, have become an essential source of information for anomaly detection~\cite{10.1145/3133956.3134015, 10.1145/3338906.3338931, FreeLog, midlog, eagerlog}. Log-based anomaly detection techniques hold significant promise for enhancing system reliability and have emerged as a research hotspot in the current field.

Existing log-based anomaly detection models can mainly be divided into unsupervised and supervised models. Unsupervised models~\cite{10.1145/3133956.3134015, 10.1145/3377813.3381371, ijcai2019p658, 9240683} use sequential neural networks, such as LSTM and GRU, to learn the occurrence probabilities of log events in normal event sequences, predicting subsequent log events and identifying events that deviate from the predictions as anomalies. However, due to the lack of explicit labeling of anomaly logs, the detection capability of these models is somewhat limited~\cite{9401970}. In contrast, supervised models~\cite{8854736, 10.1145/3338906.3338931} construct classification models to identify anomalous logs, typically demonstrating higher detection performance. However, their effectiveness largely depends on the availability of a large number of labeled logs. In real-world software systems, due to the large number of logs and the fact that anomaly logs are often buried among a large number of normal logs, obtaining accurate labels is a scarce and complex task~\cite{10.1145/3534678.3539106}. To address this issue, existing research mainly follows two strategies. The first strategy employs clustering methods to derive fuzzy log labels~\cite{9401970}, but still relies on a large number of labeled anomaly logs, which is often difficult to achieve in many practical scenarios. The second strategy introduces active learning techniques for the acquisition of online labels~\cite{10248257, 10.1145/3534678.3539106}. However, the manual online labeling process is complex and labor intensive, as each anomaly needs to be individually evaluated for its validity. Additionally, models based on online labeling also face a cold-start problem, where their performance heavily depends on the gradual accumulation of labels. Therefore, the deployment of supervised models in new software systems is extremely challenging.

To address the above issues, several cross-system log-based anomaly detection methods have been proposed, such as LogTransfer~\cite{9251092}, LogTAD~\cite{10.1145/3459637.3482209} and MetaLog~\cite{10.1145/3597503.3639205}. These methods leverage abundant labeled logs from the source system and partially labeled logs from the target system to construct anomaly detection models. LogTransfer and LogTAD rely on simple transfer learning models, where the source and target systems share part of the neural network architecture. However, existing studies have shown that transfer learning methods can only guarantee their performance under specific assumptions. When there is a significant distribution discrepancy between the source and target domains, transfer learning may encounter substantial difficulties~\cite{7542175, ijcai2022p496, NIPS2014_375c7134, DBLP:journals/corr/abs-1802-03596}. As a result, the generalization capability of these methods is heavily constrained. In contrast, MetaLog employs meta-learning to enhance generalization ability. Meta-learning involves external optimization, enabling MetaLog to handle broader meta-representations beyond just model parameters~\cite{pmlr-v70-finn17a, 9428530}. Existing research has shown that compared to transfer learning, meta-learning can achieve comparable generalization results with fewer data~\cite{NEURIPS2020_cfee3986, gu-etal-2018-meta}. Therefore, in the latest cross-system studies, MetaLog demonstrates strong cross-system performance with few-label target logs.
However, MetaLog still requires labeled logs from the target system to achieve its anomaly detection performance, because MetaLog adopts an optimization-based meta-learning paradigm that relies on a labeled “support set” of both normal and anomalous logs to compute supervised losses for fine-tuning its initialization; its classification mechanism further depends on these labeled logs to construct class prototypes and calibrate decision boundaries in the embedding space. Without any labeled logs in the target system, there is no supervision signal for adaptation, no way to form reliable class prototypes, and no means to perform cross-system embedding alignment, making zero-label log-based anomaly detection infeasible. As labeling system logs often requires huge human efforts~\cite{hilog, aclog, afalog} and in some scenarios such as system cold-start when labeled logs are not available, we aim to reduce the required labeled logs to zero and study a valuable yet underexplored setting: zero-label generalizable cross-system log-based anomaly detection.

Our goal is to generalize the model capabilities trained on labeled logs from a mature system (source system) to effectively detect anomalies in the zero-labeled logs of a new system (target system). However, achieving this is not easy. First, how can we learn discriminative feature representations on a completely unlabeled target system? Due to the absence of in-domain anomaly and normal labels, we are unable to directly fine-tune our model using supervised losses or calibrate the decision boundary between normal and anomalous events; as a result, we cannot align the model’s feature space with the true data distribution of the target system and must rely solely on the source domain label information, which may not capture the target’s unique patterns of normality and deviation. Second, how can we overcome the substantial distributional and semantic shifts between software systems to achieve robust zero-label generalization? Without any corrective feedback from the target domain, features learned on the source system may suffer from semantic mismatch, where identical log patterns carry different meanings across systems, and distribution drift, where the source feature manifold fails to cover the target’s data space, making it difficult to dynamically adapt network parameters to the target’s feature distribution.

Facing these two challenges, we propose ZeroLog, a system-agnostic representation meta-learning method that leverages both the anomalous classification features of labeled logs from the source domain and domain-invariant features between the source and target domains. Unlike MetaLog, which depends on labeled target logs for domain-specific adaptation, ZeroLog instead combines unsupervised adversarial domain adaptation with meta-learning to learn and transfer truly system-agnostic feature representations without any target domain annotations. Specifically, to address the first challenge, we use unsupervised domain adaptation to perform adversarial training between the source and target domains, enabling the learning of system-agnostic general feature representations. To tackle the second challenge, we apply meta-learning to transfer the learned representations to the target system. Based on these two methods, ZeroLog achieves robust cross-system performance under zero-label conditions. We evaluate the performance of ZeroLog on three public log datasets from different systems (HDFS, BGL and OpenStack). Results show that under zero-label conditions, ZeroLog achieves an F1-score exceeding 80$\%$, performing comparably to the latest cross-system log-based anomaly detection methods trained with labeled logs. Furthermore, ZeroLog significantly outperforms cross-system methods in the zero-label setting for the target system. 

In summary, the contributions of this paper are as follows:
\begin{itemize}
    \item{We study a valuable yet underexplored setting: zero-label generalizable cross-system log-based anomaly detection, and propose ZeroLog, a novel method that eliminates the need for labeled target system logs, enabling cross-system log-based anomaly detection under zero-label conditions.}
    \item{To achieve zero-label generalization, we design a system-agnostic representation meta-learning method, which effectively utilizes the anomaly classification features of labeled logs in the source domain and domain-invariant features between the source and target domains, for cross-system log-based anomaly detection.}
    \item{Evaluation results on three public log datasets demonstrate the significant effectiveness of our method.}
\end{itemize}

\section{Preliminary}
\label{sec2}
\subsection{System Logs and Log Events}
\begin{figure}[h!]
\centering
\includegraphics[width=3.5in]{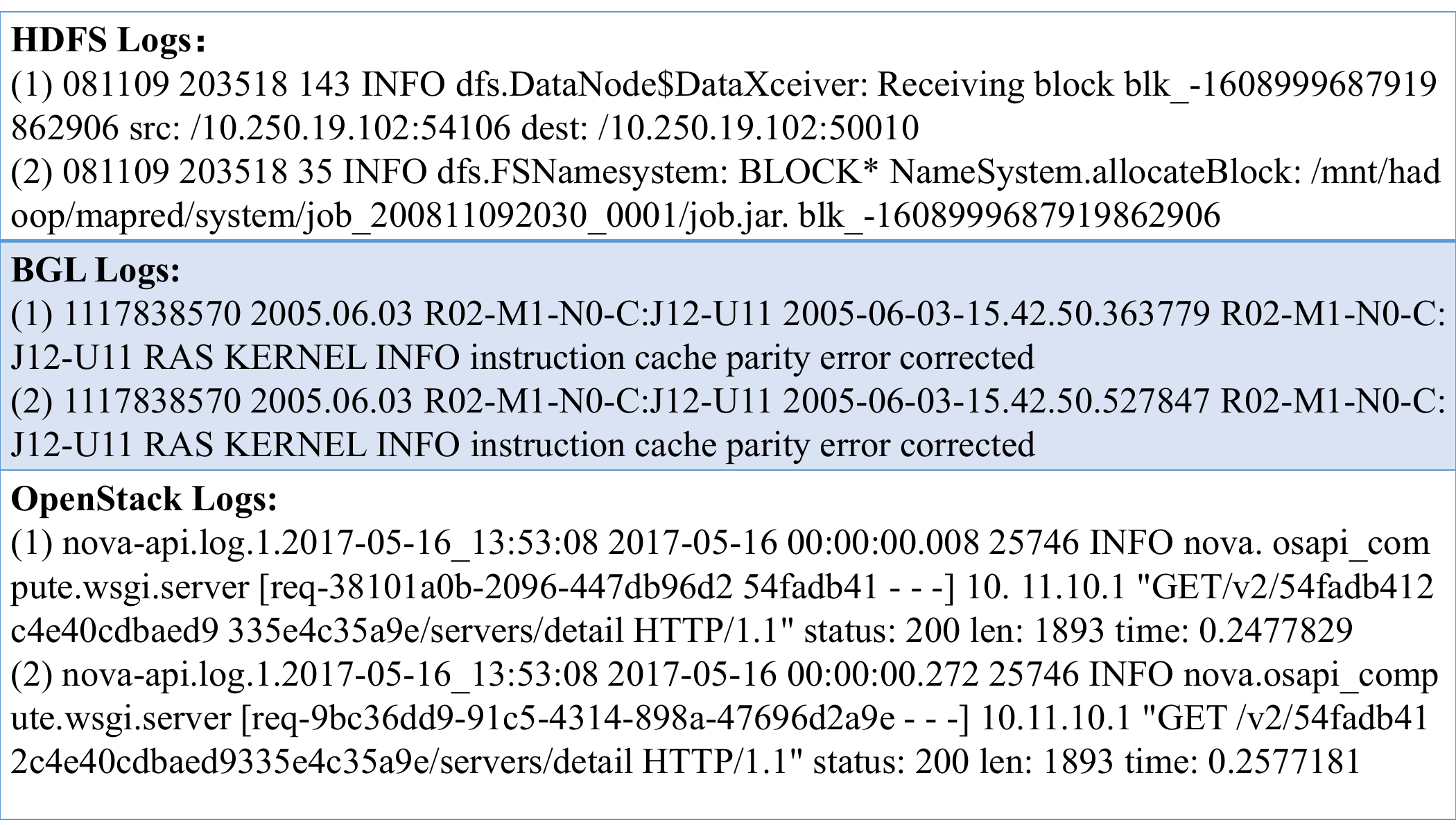}
\vspace{-0.5cm}
\caption{HDFS, BGL and OpenStack Logs.}
\label{fig_1}
\end{figure}

Software systems periodically record their operational status (such as parameters, execution state, events, etc.) in the form of text message sequences within logs. A log sequence consists of multiple log entries arranged chronologically.  An event is an abstraction of a print statement in source code, which manifests itself in logs with different embedded parameter values in different executions—represented as a set of invariant keywords and parameters. An event can be used to summarize multiple log entries. An event sequence consists of a sequence of log events in one to one correspondence with log entries in a log sequence. As shown in Figure \hyperref[fig_1]{1}, logs from HDFS, BGL and OpenStack exhibit significant variations in data structure and semantics. HDFS logs record the operations on the distributed file system, file access, node status, error information, and more. BGL logs record the hardware status, task scheduling, and execution of parallel computing tasks in supercomputing systems. OpenStack logs capture operations, events, and errors of various components in a cloud platform. They exhibit significant differences in terms of field formats (e.g., timestamp precision), log template granularity (operation-level vs. system-level), semantic space (storage operations vs. computational tasks vs. cloud services), and data scale. Existing log-based anomaly detection methods typically consist of two main steps: log processing and model training. The task of log processing is to parse the system logs and apply necessary transformations to the parsed log events, such as semantic embedding. Model training involves constructing anomaly detection models based on the features.

\vspace{-0.2cm}
\vspace{-0.3cm}

\subsection{Scenario Specification for Zero‑Label Cross‑System Log‑Based Anomaly Detection}
In the zero‑label scenario, a new system at cold start has no labeled logs, rendering all existing cross‑system log-based anomaly detection methods—including state‑of‑the‑art few‑label approaches—ineffective. The fundamental technical challenge distinguishing zero-label from few-label settings lies in the complete absence of supervision from the target system and the inability to calibrate anomaly detection models for domain-specific patterns. Unlike few-label settings, where a small set of labeled samples can guide adaptation to domain shifts, zero-label generalization must rely solely on labeled data from the source system and unlabeled data from the target system, making effective knowledge transfer and generalization across heterogeneous log distributions significantly more difficult. Our goal is to address the task of cross-system log-based anomaly detection under the condition of zero-label logs in the target system. We refer to this task as zero-label generalizable cross-system log-based anomaly detection. Specifically, this task presents two key challenges. First, the significant data differences between different systems pose a considerable challenge. Second, we can't get any labeling information from the target system. This paper aims to tackle these issues through zero-label generalization. In the zero-label generalization scenario, the model is trained using labeled logs from the source system and unlabeled logs from the target system, followed by anomaly detection on the target logs.

\begin{figure*}[h!]
\centering
\includegraphics[width=7in]{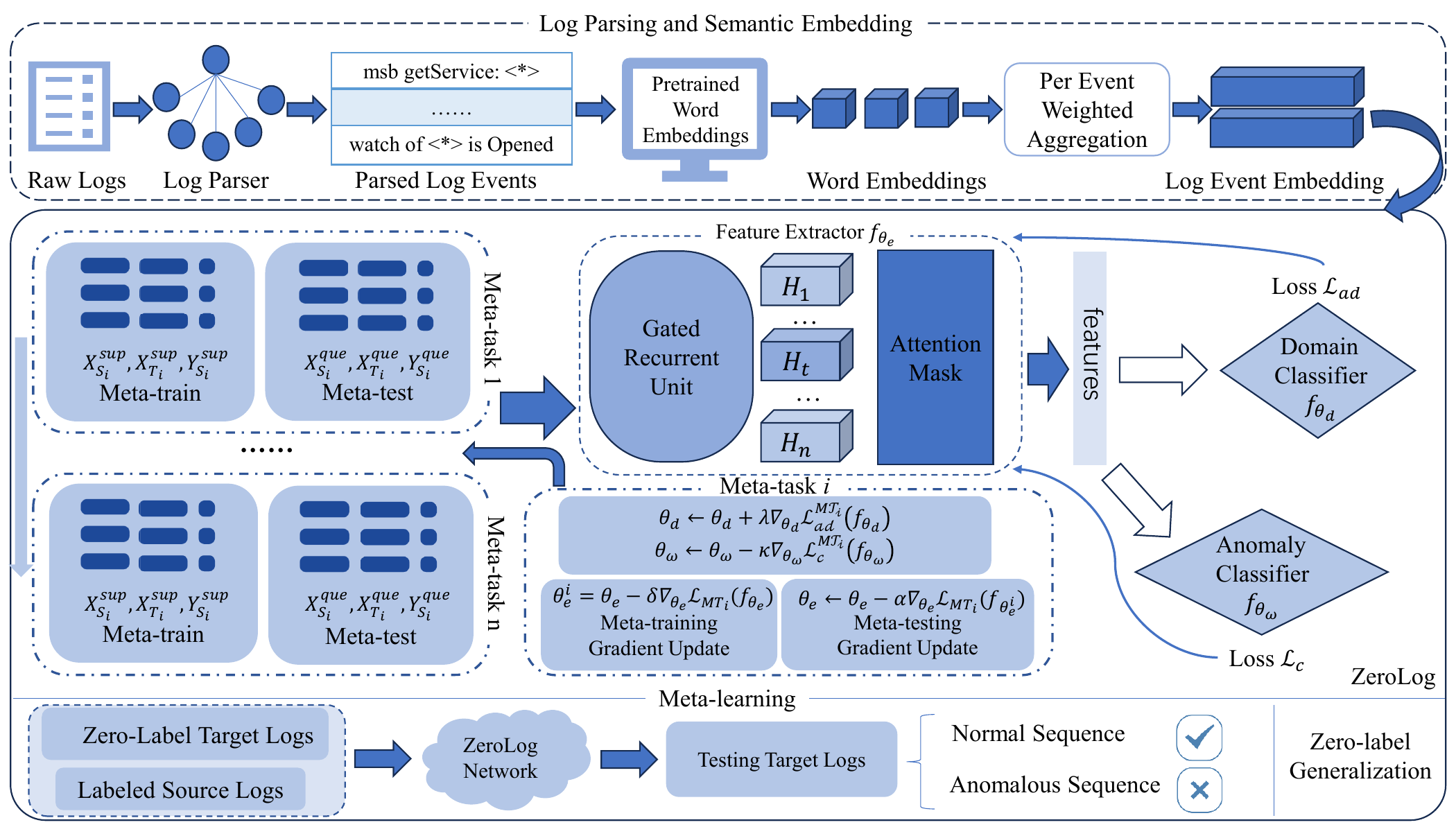}
\vspace{-0.2cm}
\caption{The proposed zero-label generalizable cross-system log-based anomaly detection pipeline for ZeroLog. ZeroLog pipeline comprises two stages: log processing and system-agnostic representation meta-learning. In the log processing stage, raw logs are parsed into log events. Subsequently, the parsed log events from different systems undergo a globally consistent semantic embedding process to generate log event embeddings. Finally, in the meta-learning stage, the ZeroLog network leverages the sequences of log event embeddings to predict whether the log sequences are normal or anomalous.}
\label{fig_2}
\end{figure*}

\section{Method}
\label{sec3}
\subsection{Overview}
Cross-system log-based anomaly detection is critical to maintaining the reliability of industrial systems. Existing methods, such as LogTAD, LogTransfer and MetaLog, rely on labeled logs from the target system to achieve cross-system performance. Even the most effective method, MetaLog, can only generalize across systems when a limited amount of labeled data from the target system is available, failing to achieve effective generalization under a zero-label setting. Under zero-label conditions, the feature representations of the target system are entirely unknown, making zero-label cross-system log-based anomaly detection particularly challenging. To tackle this challenge, we propose ZeroLog, a novel cross-system log-based anomaly detection method, which consists of three key modules: Log Embedding, System-Agnostic Representation Meta-Learning and ZeroLog Network. By leveraging the unique advantages they provide, ZeroLog is able to learn system-agnostic general feature representations from the system's logs, and generalize the representations to a specific target system through a meta-learning process.

ZeroLog begins by employing the classic log parsing technique Drain~\cite{8029742} to process unstructured raw logs from various software systems and extract log events. Compared to traditional index-based methods, semantic embeddings have been shown to provide more informative representations. Therefore, we generate semantic embeddings for each log event using methods consistent with prior studies~\cite{9251092, 10.1145/3459637.3482209, 9401970, 10.1145/3338906.3338931}. To account for the cross-system nature of log-based anomaly detection, we adopt the semantic embedding approach inspired by MetaLog~\cite{10.1145/3597503.3639205}, which ensures consistency in event representations by constructing semantic embedding vectors for log events within a shared global space. Finally, these log event embeddings are fed into the ZeroLog network (see Section ~\ref{sec3.3}) for anomaly detection (refer to Figure \hyperref[fig_3]{3}). During the process, the system-agnostic representation meta-learning plays a central role, as detailed in Section ~\ref{sec3.2}. It effectively utilizes the anomaly classification features of labeled logs in the source domain, as well as domain-invariant features between the source and target domains, for cross-system log-based anomaly detection. This process facilitates the network in acquiring the necessary generalization capability between the source and target system logs. The complete workflow of the proposed ZeroLog is illustrated in Figure \hyperref[fig_2]{2}, and the overall procedure of the algorithm is presented in Algorithm \hyperref[alg]{1}.

\begin{algorithm}[!h]
    \caption{ZeroLog}
    \label{alg}
    \renewcommand{\algorithmicrequire}{\textbf{Input:}}
    \renewcommand{\algorithmicensure}{\textbf{Output:}}
    \begin{algorithmic}[1]
        \REQUIRE Source logs $D_S$, Target logs $D_T$, Task distribution $p(MT)$, Learning rates $\delta$, $\lambda$, $\kappa$, Hyperparameters $\alpha$, $\beta$, $\gamma.$  
        \ENSURE Feature extractor \( f_{\theta_e} \), Anomaly classifier \( f_{\theta_\omega} \) and Domain classifier \( f_{\theta_d}. \)    
        \STATE  Randomly initialize \( f_{\theta_e} \), \( f_{\theta_\omega} \), \( f_{\theta_d}. \)
        \WHILE{not done}
            \STATE Sample batch of tasks \( MT_i = \{ M_i^{sup}, M_i^{que} \}\sim p(MT)\).
            \STATE Evaluate $\nabla_{\theta_d}\sum_{MT_i\sim p(MT)}\mathcal{L}_{ad}^{MT_i}(f_{\theta_d})$ on $M_i^{sup}$ by using Eq. (1).
            \STATE Update the domain classifier \( f_{\theta_d} \) with gradient ascent: $\theta_d \gets \theta_d + \lambda \nabla_{\theta_d} \sum_{MT_i \sim p(MT)} \mathcal{L}_{ad}^{MT_i}(f_{\theta_d}).$
            \STATE Evaluate $\nabla_{\theta_\omega}\sum_{MT_i\sim p(MT)}\mathcal{L}_{c}^{MT_i}(f_{\theta_\omega})$ on $M_i^{sup}$ by using Eq. (2).
            \STATE Update the anomaly classifier \( f_{\theta_\omega} \) with gradient ascent: $\theta_\omega \gets \theta_\omega - \kappa \nabla_{\theta_\omega} \sum_{MT_i \sim p(MT)} \mathcal{L}_{c}^{MT_i}(f_{\theta_\omega}).$
            \FOR{all $MT_i$}
                \STATE Evaluate $\nabla_{\theta_e}\mathcal{L}_{MT_i}(f_{\theta_e})$ on $M_i^{sup}$ by using Eq. (3).
                \STATE Compute adapted parameters with gradient descent: \( \theta_e^i = \theta_e - \delta \nabla_{\theta_e} \mathcal{L}_{MT_i}(f_{\theta_e}). \)
            \ENDFOR
            \STATE Evaluate $\nabla_{\theta_e} \sum_{MT_i \sim p(MT)} \mathcal{L}_{MT_i}(f_{\theta_e^i})$ on $M_i^{que}$ by using Eq. (4).
            \STATE Update the feature extractor \( f_{\theta_e} \) with gradient descent: $\theta_e \gets \theta_e - \alpha \nabla_{\theta_e} \sum_{MT_i \sim p(MT)} \mathcal{L}_{MT_i}(f_{\theta_e^i}).$
        \ENDWHILE
        \RETURN Final ZeroLog network parameters.
    \end{algorithmic}
\end{algorithm}

\subsection{System-Agnostic Representation Meta-Learning}
\label{sec3.2}
To achieve zero-label generalization, we use unsupervised domain adaptation (UDA) to perform adversarial training between the source and target domains, enabling the learning of system-agnostic feature representations. UDA addresses the challenge of limited model generalization caused by distribution discrepancies between the source and target domains. UDA methods based on adversarial learning employ adversarial training mechanisms to extract domain-invariant features, achieving alignment between the source and target domain distributions. These methods typically comprise a feature extractor and a domain classifier. The feature extractor is optimized to generate features indistinguishable by the domain classifier, thereby mitigating domain-specific biases while retaining task-relevant information. This adversarial optimization fosters robust feature representations that generalize effectively across domains. Then, we apply meta-learning to transfer the learned representations to the target system. Meta-learning enhances a model's adaptability to new tasks by extracting shared task-relevant features and optimizing learning strategies during multi-task training. By leveraging parameter sharing, meta-learning captures commonalities across tasks while suppressing domain-specific variations. Furthermore, it optimizes the model’s initialization to enable rapid convergence on unseen tasks with minimal gradient updates. During inference, the model efficiently adapts to task-specific requirements through fast fine-tuning or direct utilization of shared representations, ensuring strong generalization and adaptability. 

The setup of ZeroLog involves two domains: the source domain \( D_S \) and the target domain \( D_T \). Formally, \( X_S \) and \( X_T \) represent the data sampled from \( D_S \) and \( D_T \), respectively, while \( Y_S \) denotes the label matrix for \( X_S \). In meta-learning, the smallest unit for gradient updates is the meta-task. A meta-task consists of two phases: the meta-training phase and the meta-testing phase. To achieve zero-label generalization, designing appropriate cross-system meta-tasks is critical for ZeroLog. We construct a cross-system meta-task \( MT_i = \{ M_i^{sup}, M_i^{que} \} \sim p(MT) \), where $p(MT)$ is task distribution, \( M_i^{sup} = \{ X_{S_i}^{sup}, X_{T_i}^{sup}, Y_{S_i}^{sup} \} \) is used for meta-training, and \( M_i^{que} = \{ X_{S_i}^{que}, X_{T_i}^{que}, Y_{S_i}^{que} \} \) is used for meta-testing. Here, \( X^{sup} \) and \( X^{que} \) are referred to as the support set and query set, respectively, with \( X_{S_i} \) and \( X_{T_i} \) sampled from the source and target logs. ZeroLog consists of three main modules: the feature extractor \( f_{\theta_e} \), the anomaly classifier \( f_{\theta_\omega} \) and the domain classifier \( f_{\theta_d} \). To handle domain transfer, we introduce a domain classifier to maximize the distinction between source and target domain features, while leveraging adversarial training to align the domains. For each meta-task, the anomaly classifier attempts to find a local optimal solution for the classification problem, while both the anomaly and domain classifier iteratively update across tasks to find the global optimal solution. In the following, we will elaborate on our method, which can be divided into three key components.

\textit{Training the Domain Classifier and Anomaly Classifier.} Given the feature extractor \( f_{\theta_e} \), we train the domain classifier \( f_{\theta_d} \) on the meta-task. The optimization is formulated as:
\begin{equation}
\max_{\theta_d} \sum_{MT_i \sim p(MT)} \mathcal{L}_{ad}^{MT_i}(M_i^{sup}; f_{\theta_d}). \label{1}
\end{equation}
Here, the adversarial loss function \( \mathcal{L}_{ad}^{MT_i}(f_{\theta_d}) \) is computed as:
\[
\mathcal{L}_{ad}^{MT_i}(f_{\theta_d}) = \log(f_{\theta_d}(f_{\theta_e}(X_{S_i}^{sup}))) + \log(1 - f_{\theta_d}(f_{\theta_e}(X_{T_i}^{sup}))).
\]
The purpose of training a domain classifier is to maximize the distinction between features from the source and target domains. Subsequently, adversarial training is conducted with the feature extractor, encouraging the feature extractor to learn domain-invariant features that are shared between the source and target domains. Then, we train the anomaly classifier \( f_{\theta_\omega} \) on the meta-task. The optimization is formulated as:
\begin{equation}
\min_{\theta_\omega} \sum_{MT_i \sim p(MT)} \mathcal{L}_c^{MT_i}(M_i^{sup}; f_{\theta_\omega}), \label{2}
\end{equation}
where \( \mathcal{L}_c^{MT_i}(f_{\theta_\omega}) \) is the Binary Cross-Entropy loss. The objective of training the anomaly classifier is to learn discriminative features for classifying normal and anomalous logs, thereby achieving robust anomaly detection performance.

\textit{Adapting to Meta-Tasks.} To train a feature extractor \( f_{\theta_e} \) that can effectively adapt to the target system's features, we learn to adapt the meta-task \( MT_i \). During the meta-training phase, the learner's parameters \( \theta_e \) become \( \theta_e^i \), which can be updated through one or more gradient descent steps: $\theta_e^i = \theta_e - \delta \nabla_{\theta_e} \mathcal{L}_{MT_i}(M_i^{sup}; f_{\theta_e})$, where \( \delta \) is the learning rate. In \( MT_i \), the goal is to achieve effective classification while performing domain adaptation. Therefore, the objective function for the current task can be written as:
\begin{equation}
\small
\mathcal{L}_{MT_i}(f_{\theta_e}) = \gamma \mathcal{L}_c^{MT_i}(X_{S_i}^{sup}, Y_{S_i}^{sup}; f_{\theta_\omega}) + \beta \mathcal{L}_{ad}^{MT_i}(X_{S_i}^{sup}, X_{T_i}^{sup}; f_{\theta_d}). 
\label{3}
\end{equation}
The first term represents the classification loss in the source domain with labeled information. The second term is the domain adversarial loss, which encourages the feature extractor \( f_{\theta_e} \) to produce domain-invariant features by aligning the domains through the domain classifier \( f_{\theta_d} \). The hyperparameters \( \beta \) and \( \gamma \) control the trade-off between adaptation and classification performance. This method integrates classification loss and adversarial loss, enabling ZeroLog to effectively generalize from the source system to the target system.

\textit{Meta-Optimization of Tasks.} After learning the adaptation parameters \( \theta_e^i \) for each task, we proceed to meta-optimize the feature extractor \( f_{\theta_e} \). The objective is to improve the performance of \( \theta_e^i \) on the query set. Using the parameters \( \theta_e^i \) obtained from meta-training, we compute the corresponding loss term \( \mathcal{L}_{MT_i}(M_i^{que}; f_{\theta_e^i}) \). The meta-objective function can be expressed as:
\begin{equation}
\min_{\theta_e} \sum_{MT_i \sim p(MT)} \mathcal{L}_{MT_i}(M_i^{que}; f_{\theta_e^i}). \label{4}
\end{equation}
We perform meta-optimization via gradient descent as follows:
\[
\theta_e \gets \theta_e - \alpha \nabla_{\theta_e} \sum_{MT_i \sim p(MT)} \mathcal{L}_{MT_i}(M_i^{que}; f_{\theta_e^i}),
\]
where \( \alpha \) is the meta-step size. During meta-optimization, we apply domain adversarial training to the query set through the domain classifier, which ensures domain alignment. Therefore, the goal of the meta-optimization process is to learn a general feature extractor that can quickly adapt to new tasks involving both classification and domain adaptation.

Overall, the goal of the ZeroLog algorithm is to teach the ZeroLog network to generalize from source system logs to target system logs through carefully designed tasks. During the meta-training phase, the parameters of the feature extractor are updated using gradients from the source. In the meta-testing phase, the corresponding loss terms are computed based on the meta-trained parameters. Once the meta-testing loss terms are obtained, the meta-learning process optimizes the feature extractor. Ultimately, optimal meta-parameters are achieved through training on the meta-tasks. The total loss for ZeroLog can be expressed as a min-max optimization problem:
\[
\min_{\theta_e} \max_{\theta_d} \sum_{MT_i \sim p(MT)} \left( \gamma  \mathcal{L}_c^{MT_i}(f_{\theta_e}, f_{\theta_\omega}) + \beta  \mathcal{L}_{ad}^{MT_i}(f_{\theta_e}, f_{\theta_d}) \right).
\]
This integrated optimization strategy takes into account information from both the source and target systems, providing sufficient cues for the ZeroLog network to solve the zero-label cross-system log-based anomaly detection task.

\subsection{ZeroLog Network}
\label{sec3.3}
\begin{figure}[h!]
\centering
\includegraphics[width=3.5in]{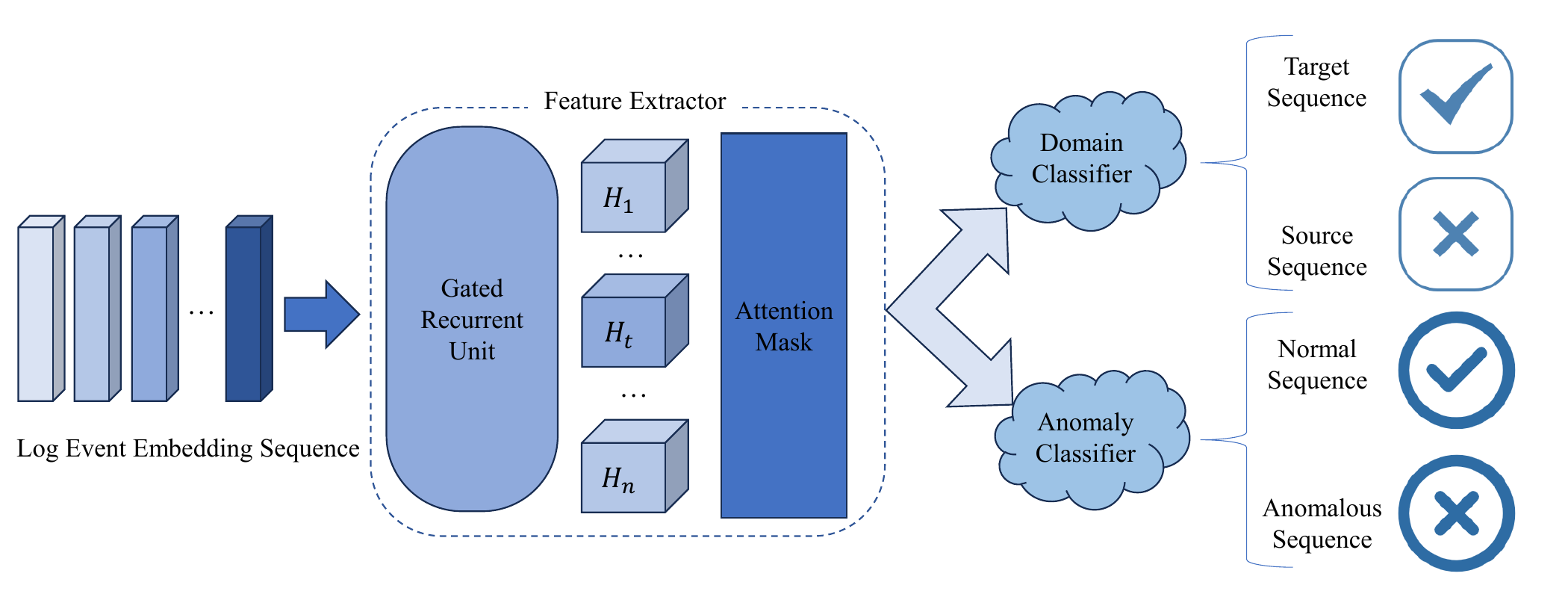}
\vspace{-0.5cm}
\caption{The ZeroLog network architecture.}
\label{fig_3}
\end{figure}

In this section, we provide a clear explanation of the ZeroLog network, with the detailed design shown in Figure \hyperref[fig_3]{3}. Like Metalog~\cite{10.1145/3597503.3639205}, the feature extractor consists of two interconnected modules: the Gated Recurrent Unit (GRU) and the attention mask layer. Given a sequence of log event embeddings \( \{V_{e_1}, V_{e_2}, \dots, V_{e_n}\} \), where \( e_t \) represents the \( t \)-th log event in the sequence, the GRU module maintains a hidden state \( H_t \) at each time step, enabling the network to retain long-term information from the input log event sequence. For time step \( t \), the attention module takes the hidden states \( \{H_1, \dots, H_t\} \) from GRU as input and utilizes adaptive self-attention to fuse the information. The final representation of the log sequence combines all the previous information. The output of \( f_{\theta_e} \) is input to the anomaly classifier, which generates an anomaly probability. Additionally, after the feature extractor processes the log features, the resulting feature tensor is input into the domain classifier, which processes it and ultimately outputs a classification result.

\vspace{-0.5cm}
\vspace{-0.3cm}

\begin{table*}[h!]
\centering
\setlength{\belowcaptionskip}{0.1cm}
\caption{Statistics of the Datasets.}
\begin{tabular}{|c|c|c|c|c|c|c|}
\hline
\textbf{Dataset} & \textbf{Amount of Lines} & \textbf{Amount of log sessions} & \textbf{Total Normal} & \textbf{Total Anomalous} & \textbf{Used Normal} & \textbf{Used Anomalous} \\ \hline
HDFS~\cite{10.1145/1629575.1629587} & 11,175,629 & 575,061  & 558,223 & 16,838 & 17,406 & 16,838 \\ \hline
BGL~\cite{4273008} & 4,747,963 & 85,576  & 49,273 & 36,303 & 36,401 & 36,303 \\ \hline
OpenStack~\cite{10.1145/3133956.3134015} & 207,820 & 3,367  & 2,490 & 877 & 2,490 & 877 \\ \hline
\end{tabular}
\label{tab5}	
\end{table*}

\begin{table*}[h!]
\centering
\setlength{\belowcaptionskip}{0.1cm}
\caption{Zero-label generalization experiments across different domains.}
\label{tab_all}
\resizebox{\textwidth}{!}{%
\begin{tabular}{@{}lcccccccccccc@{}}
\toprule
\multicolumn{1}{c}{\multirow{2}{*}{\textbf{Method}}} & \multicolumn{3}{c}{\textbf{HDFS to BGL}} & \multicolumn{3}{c}{\textbf{BGL to HDFS}} & \multicolumn{3}{c}{\textbf{OpenStack to HDFS}} & \multicolumn{3}{c}{\textbf{OpenStack to BGL}} \\
\cmidrule(lr){2-4} \cmidrule(lr){5-7} \cmidrule(lr){8-10} \cmidrule(lr){11-13}
 & \textbf{Precision} & \textbf{Recall} & \textbf{F1-score} & \textbf{Precision} & \textbf{Recall} & \textbf{F1-score} & \textbf{Precision} & \textbf{Recall} & \textbf{F1-score} & \textbf{Precision} & \textbf{Recall} & \textbf{F1-score} \\
\midrule
PLELog (a1) ~\cite{9401970} & 82.10 & 67.42 & 74.04 & 65.86 & 71.11 & 68.38 & 65.86 & 71.11 & 68.38 & 82.10 & 67.42 & 74.04 \\
LogRobust (a2) ~\cite{10.1145/3338906.3338931} & 94.60 & 72.95 & 82.38 & 100.00 & 62.30 & 76.77 & 100.00 & 62.30 & 76.77 & 94.60 & 72.95 & \underline{82.38} \\ \hline
PLELog (b1) ~\cite{9401970} & 94.88 & 89.62 & \underline{92.18} & 96.30& 83.81& \underline{89.62} & 96.30& 83.81& \underline{89.62} & 94.88 & 89.62 & \underline{92.18} \\
LogRobust (b2) ~\cite{10.1145/3338906.3338931} & 97.52 & 91.27 & \underline{94.29} & 82.54& 99.20& \underline{90.11} & 82.54& 99.20& \underline{90.11} & 97.52 & 91.27 & \underline{94.29} \\ \hline
LogTAD (c1) ~\cite{10.1145/3459637.3482209} & 78.01 & 68.51 & 72.95 & 78.80 & 71.22 & 74.82 & 71.89 & 65.51 & 68.55 & 70.72 & 65.51 & 68.02 \\
LogTransfer (c2) ~\cite{9251092} & 74.42 & 76.73 & 75.56 & 100.00 & 43.30 & 60.43 & 73.87 & 62.30 & 67.59 & 68.43 & 71.32 & 69.85 \\ 
LogDLR (c3) ~\cite{10910212}  & 79.25 & 72.56 & 75.76 & 77.78 & 70.05 & 73.71 & 73.65 & 69.80 & 71.67 & 69.58 & 64.33 & 66.85 \\ 
\hline
MetaLog (d1) ~\cite{10.1145/3597503.3639205} & 96.89 & 89.28 & \underline{92.93} & 89.29 & 74.98 & \underline{81.51} & 96.67 & 62.42 & 75.86 & 99.83 & 70.09 & \underline{82.36} \\
MetaLog (d2) ~\cite{10.1145/3597503.3639205} & 64.80 & 3.62 & 6.86 & 99.93 & 28.09 & 43.85 & 97.46 & 19.21 & 32.09 & 100.00 & 1.70 & 3.34 \\
MetaLog (d3) ~\cite{10.1145/3597503.3639205} & 99.85 & 16.10 & 27.73 & 70.63 & 12.14 & 20.72 & 100.00 & 19.15 & 32.15 & 100.00 & 0.70 & 1.39 \\  \hline
DeepLog (e1) ~\cite{10.1145/3133956.3134015} & 66.13 & 48.79 & 56.16 & 53.96 & 34.07 & 41.77 & 53.96 & 34.07 & 41.77 & 66.13 & 48.79 & 56.16 \\ \hline
PLELog (f1) ~\cite{9401970} & 38.80 & 99.87 & 55.89 & 1.69 & 92.85 & 3.32 & 4.33 & 53.47 & 8.01 & 54.65 & 43.04 & 48.16 \\
LogRobust (f2) ~\cite{10.1145/3338906.3338931} & 39.08 & 93.67 & 55.15 & 2.25 & 62.12 & 4.35 & 0.63 & 60.81 & 1.25 & 34.34 & 57.39 & 42.97 \\
NeuralLog (f3) ~\cite{9678773} & 57.23 & 52.79 & 54.38 & 33.13 & 58.04 & 42.06 & 3.33 & 42.99 & 6.17 & 14.76 & 81.45 & 24.99 \\
MetaLog (f4) ~\cite{10.1145/3597503.3639205} & 29.43 & 0.45 & 0.89 & 2.90 & 80.92 & 5.61 & 2.90 & 80.92 & 5.61 & 29.43 & 0.45 & 0.89 \\ \hline
Ours ZeroLog & 83.10 & 89.13 & \textbf{86.01} & 82.70 & 78.61 & \textbf{80.61} & 75.73 & 85.24 & \textbf{80.21} & 74.44 & 80.92 & \textbf{77.55} \\
\bottomrule
\end{tabular}%
}
\end{table*}

\section{Experiments}
\label{sec4}
\subsection{Experimental Setup}
\label{sec4.1}
\textit{Datasets.} We conducted systematic experiments on three publicly available log datasets: HDFS~\cite{10.1145/1629575.1629587}, BGL~\cite{4273008} and OpenStack~\cite{10.1145/3133956.3134015}. The statistics for these three datasets are summarized in Table \hyperref[tab5]{1}. In the zero-label setting, we selected four cross-system dataset combinations (HDFS to BGL, BGL to HDFS, OpenStack to HDFS and OpenStack to BGL) to validate our method. In these experiments, the full labeled logs from the source system were used to train the ZeroLog network, while the logs from the target system remained unlabeled. For the four experimental setups, we followed the code provided in~\cite{9401970} and used Drain~\cite{8029742} to parse the log events and organize the log sessions. This preprocessing ensured that the structures of all datasets were consistent with previous research methods, thus enabling a fair comparison.

Specifically, in the HDFS to BGL experiment, all anomalous logs from the HDFS dataset and an equal number of normal logs, along with the unlabeled logs from the BGL dataset (with the same setup), were used for training. This setup simulates the real-world challenge of a target system with scarce anomaly labels. In the BGL to HDFS experiment, the same settings are followed. For the OpenStack to HDFS and BGL experiments, since the OpenStack dataset is much smaller than the HDFS and BGL datasets, we used all the logs from the OpenStack dataset, along with an equal number of normal and anomalous unlabeled logs from the HDFS and BGL datasets for the training phase. The exact amount of data used can be seen in Table \hyperref[tab5]{1}.

\textit{Baselines.} To perform zero-label generalization tasks and ensure a fair comparison with our method, we adopted various baseline methods and experimental setups. Due to the significant differences in the sizes of the three datasets, we used different proportions of data for the experiments.

Supervised baseline. In table \hyperref[tab_all]{2}, block (a) and (b) presents the performance of the semi-supervised method PLELog~\cite{9401970} and the fully supervised baseline LogRobust~\cite{10.1145/3338906.3338931}. When BGL is the target system, the method of block (a) are trained on 30$\%$ of normal logs and only 1$\%$ of anomalous logs from the BGL dataset. The method of block (b) are trained on 30$\%$ of total logs from the BGL dataset. Similarly, when HDFS is the target system, the method of block (a) are trained on 10$\%$ of normal logs and 1$\%$ of anomalous logs from the HDFS dataset. The method of block (b) are trained on 10$\%$ of total logs from the HDFS dataset. Block (a) evaluates the performance of methods trained solely on the target dataset under a scenario where anomaly labels are scarce. Block (b) examines the capability of methods trained on target datasets with fully annotated (100$\%$) anomaly labels.
  
Cross-system baseline. Block (c) highlights three transfer learning baseline methods, LogTAD~\cite{10.1145/3459637.3482209}, LogTransfer~\cite{9251092} and LogDLR~\cite{10910212}. In HDFS to BGL generalization, the method of (c1) and (c2) are trained on 30$\%$ of normal logs and 1$\%$ of anomalous logs from the BGL dataset and 30$\%$ of logs from the HDFS dataset. In BGL to HDFS generalization, (c1) and (c2) presents results for these methods trained on 10$\%$ of normal logs and 1$\%$ of anomalous logs from the HDFS dataset and all logs from the BGL dataset. Similarly, in OpenStack to HDFS generalization, (c1) and (c2) reports results for methods trained on 10$\%$ of normal logs and 1$\%$ of anomalous logs from the HDFS dataset and all logs from the OpenStack dataset. In OpenStack to BGL generalization, (c1) and (c2) follows the same setup for the BGL and OpenStack dataset. In addition, (c3) randomly selects 100,000 and 10,000 normal logs from the source system and target system respectively for training. Moreover, block (d) showcases a baseline method based on meta-learning, MetaLog~\cite{10.1145/3597503.3639205}, where (d1) shares the same source and target data configurations as LogTAD~\cite{10.1145/3459637.3482209} and LogTransfer~\cite{9251092}. In the (d2) setting, we build on the above experimental configuration but remove all anomaly labels, using only the normal labels under the same settings to perform cross‑system log-based anomaly detection. In the (d3) setting, based on the ZeroLog experimental configuration, we remove all anomaly labels and, under the same settings, perform cross-system log-based anomaly detection using only normal labels. Blocks (c) and (d) evaluate cross‐system baseline methods, based on prior transfer‐learning and meta‐learning approaches, that leverage partially labeled logs from the target system.
  
Unsupervised and zero-label baseline. Block (e) presents the unsupervised baseline method DeepLog~\cite{10.1145/3133956.3134015}, which is trained exclusively on normal labels from the target dataset, maintaining consistency with other baseline methods. Block (f) evaluates the performance of PLELog~\cite{9401970}, LogRobust~\cite{10.1145/3338906.3338931}, NeuralLog~\cite{9678773} and MetaLog~\cite{10.1145/3597503.3639205} under the zero-label setup. PLELog, LogRobust and NeuralLog method are trained solely on subsets of the source datasets (all logs of BGL and OpenStack, 30$\%$ of logs of HDFS) and are directly tested on subsets of the target datasets. In the (f4) seething, the model is trained on the fully labeled logs of the two non‑target systems and then evaluated directly on the target system. Specifically, when HDFS is the target system, we train on 30\% of the BGL logs and all of the OpenStack logs, and then test on HDFS; when BGL is the target system, we train on 10\% of the HDFS logs and all of the OpenStack logs, and then test on BGL. Blocks (e) and (f) assess the performance of existing methods in the zero‑label setting for the target system.

Overall, existing methods can achieve satisfactory cross-system log-based anomaly detection performance only when labeled data is available, and they fail to address the cold-start problem. Notably, even the state-of-the-art cross-system method MetaLog is unable to perform effective anomaly detection once anomaly labels or all labels are removed.

\textit{Evaluation Metrics.} We selected precision, recall and F1-score as evaluation metrics, defined as follows:  
\(
\text{Precision} = \frac{TP}{TP + FP}, \quad \text{Recall} = \frac{TP}{TP + FN}, \quad F_1 = \frac{2 \cdot \text{Precision} \cdot \text{Recall}}{\text{Precision} + \text{Recall}},
\)
where \( TP \), \( FP \), and \( FN \) represent true positives, false positives, and false negatives, respectively. These evaluation metrics provide a effective measure of ZeroLog's capabilities in handling anomaly detection tasks.

\textit{Implementation Details.} During the meta-learning training process, all logs were uniformly sampled and divided into log splits, with each meta-task trained on a single log split. The ZeroLog network was trained on a single NVIDIA 3090 GPU using the Adam optimizer, with a batch size of 256, a meta-parameter $\alpha$ of 1.0 and a learning rate of 3e-3. Across the four different experiments, semantic embeddings were generated as 300-dimensional input log event embeddings, following the settings in~\cite{pennington-etal-2014-glove, 9401970}.

\begin{figure*}[!h]
    \centering 

    \begin{minipage}{0.49\linewidth} 
        \centering
        \includegraphics[width=\textwidth]{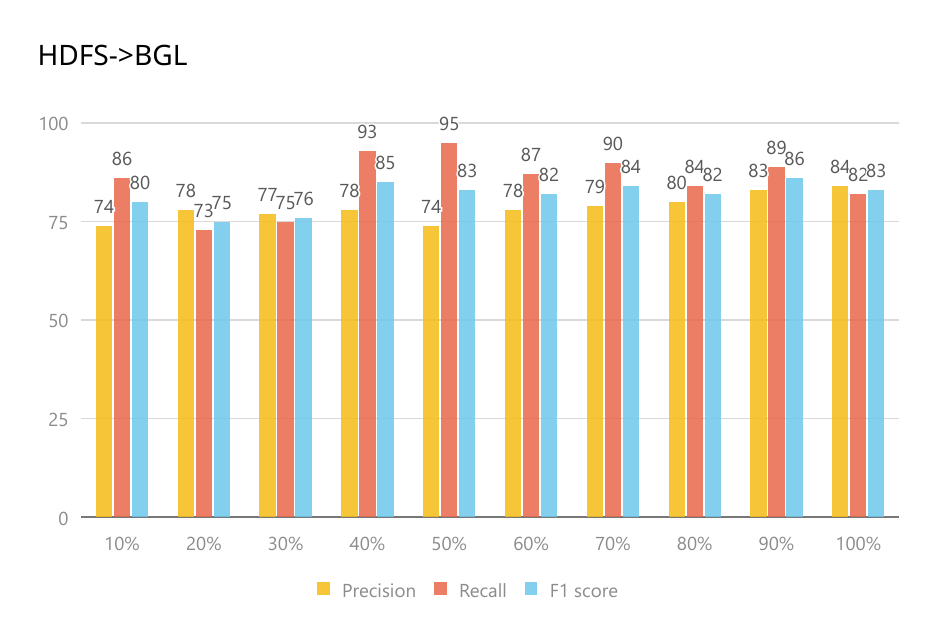}
    \end{minipage}
    \hfill
    \begin{minipage}{0.49\linewidth}
        \centering
        \includegraphics[width=\textwidth]{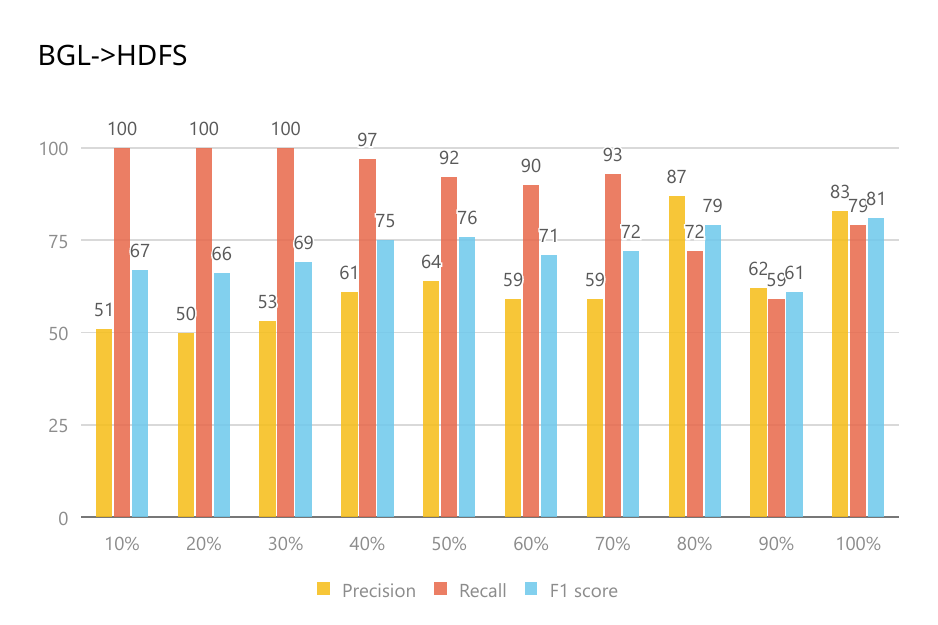}
    \end{minipage}
    \caption{Experimental study on the proportion of used source domain data.} 
    \label{fig:row1}
    
    \vspace{8pt} 

    \begin{minipage}{0.24\linewidth} 
        \centering
        \includegraphics[width=\textwidth]{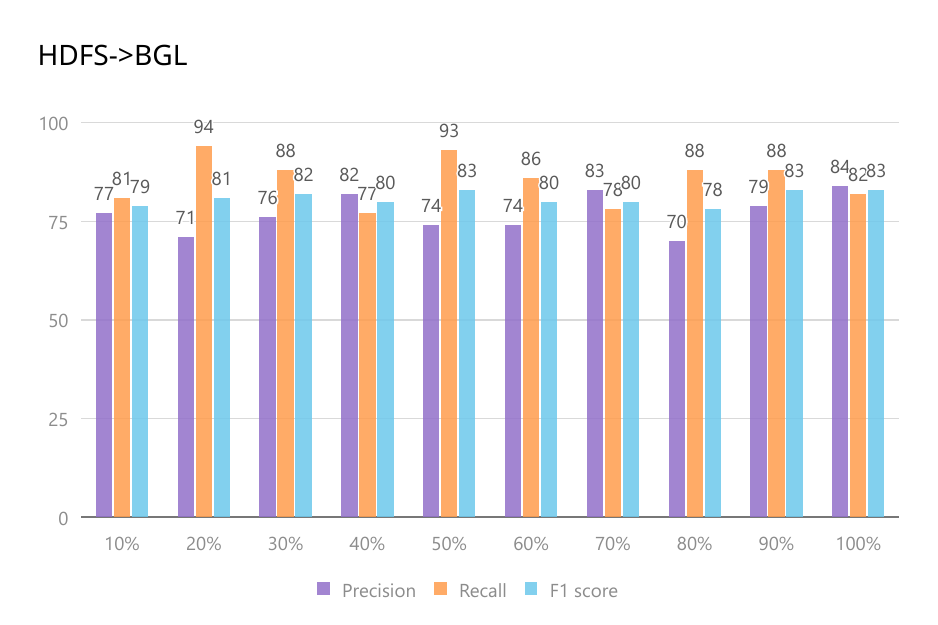}
    \end{minipage}
    \hfill
    \begin{minipage}{0.24\linewidth}
        \centering
        \includegraphics[width=\textwidth]{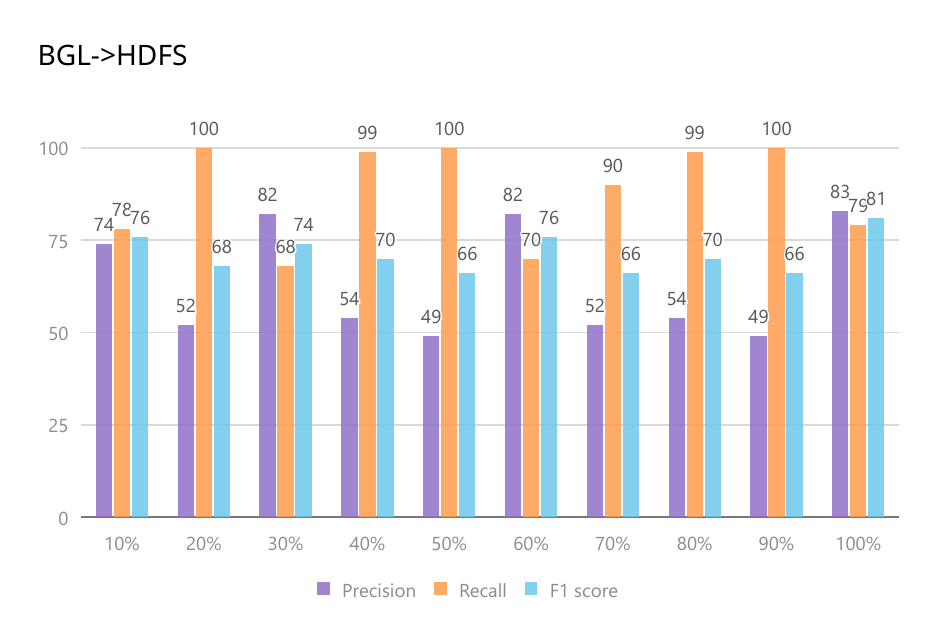}
    \end{minipage}
    \hfill
    \begin{minipage}{0.24\linewidth}
        \centering
        \includegraphics[width=\textwidth]{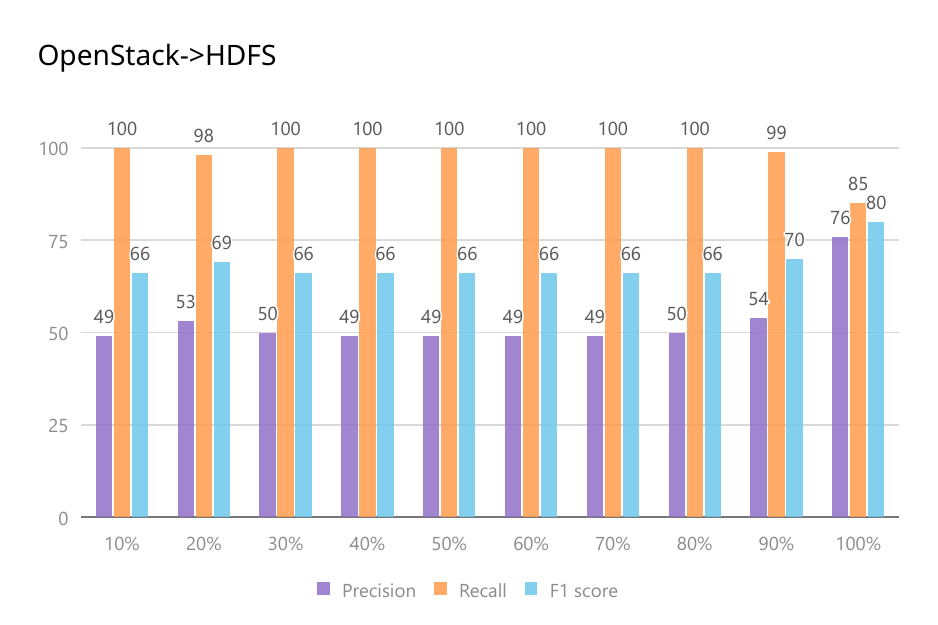}
    \end{minipage}
    \hfill
    \begin{minipage}{0.24\linewidth}
        \centering
        \includegraphics[width=\textwidth]{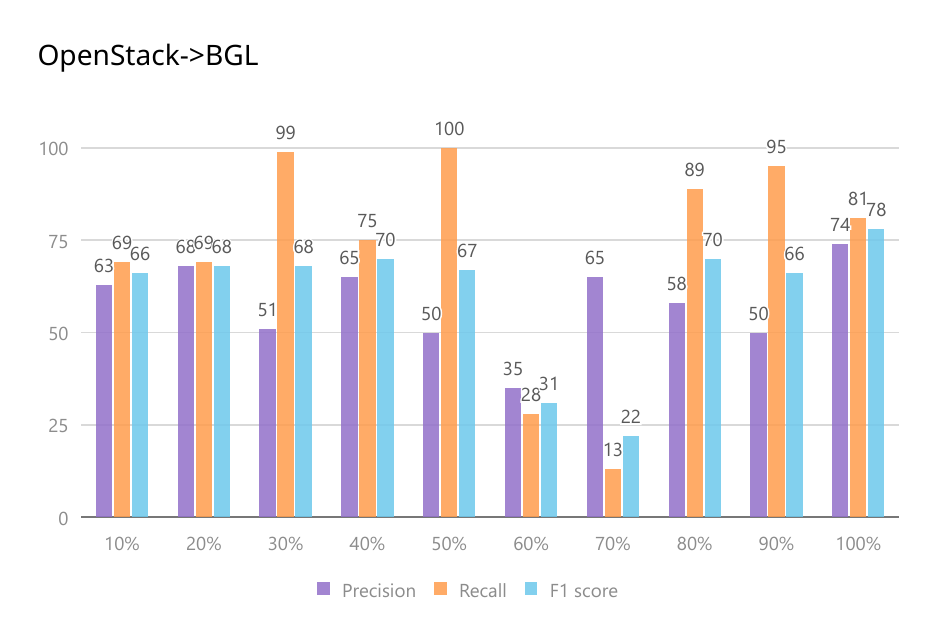}
    \end{minipage}
    \caption{Experimental study on the proportion of used target domain data.} 
    \label{fig:row2}
    
    \vspace{8pt} 

    \begin{minipage}{0.24\linewidth}
        \centering
        \includegraphics[width=\textwidth]{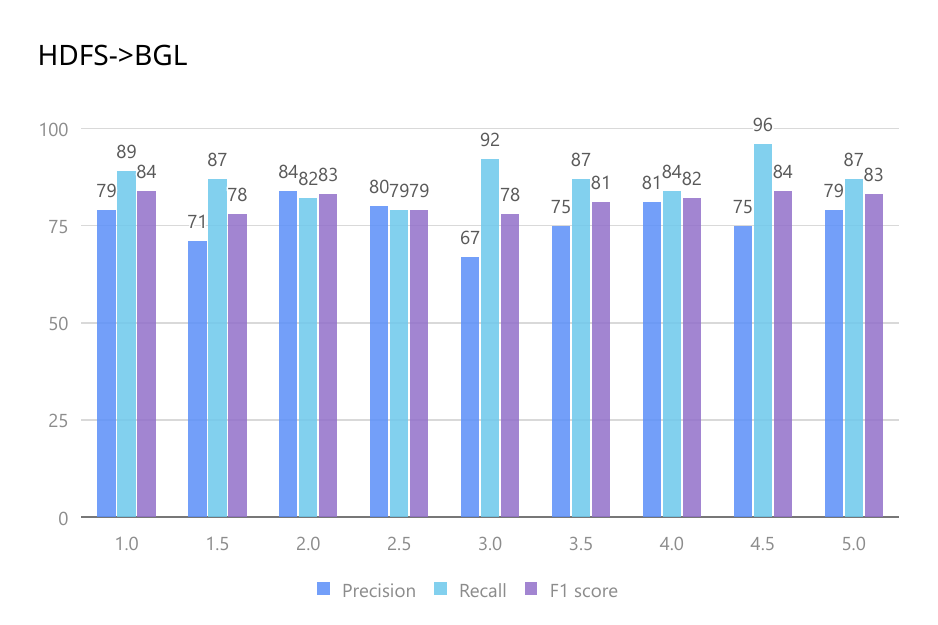}
    \end{minipage}
    \hfill
    \begin{minipage}{0.24\linewidth}
        \centering
        \includegraphics[width=\textwidth]{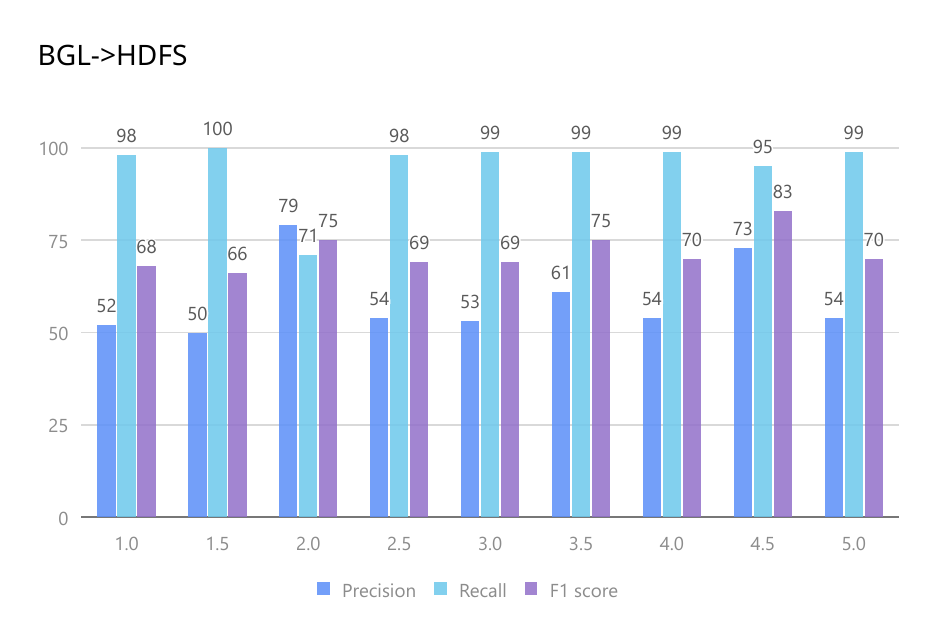}
    \end{minipage}
    \hfill
    \begin{minipage}{0.24\linewidth}
        \centering
        \includegraphics[width=\textwidth]{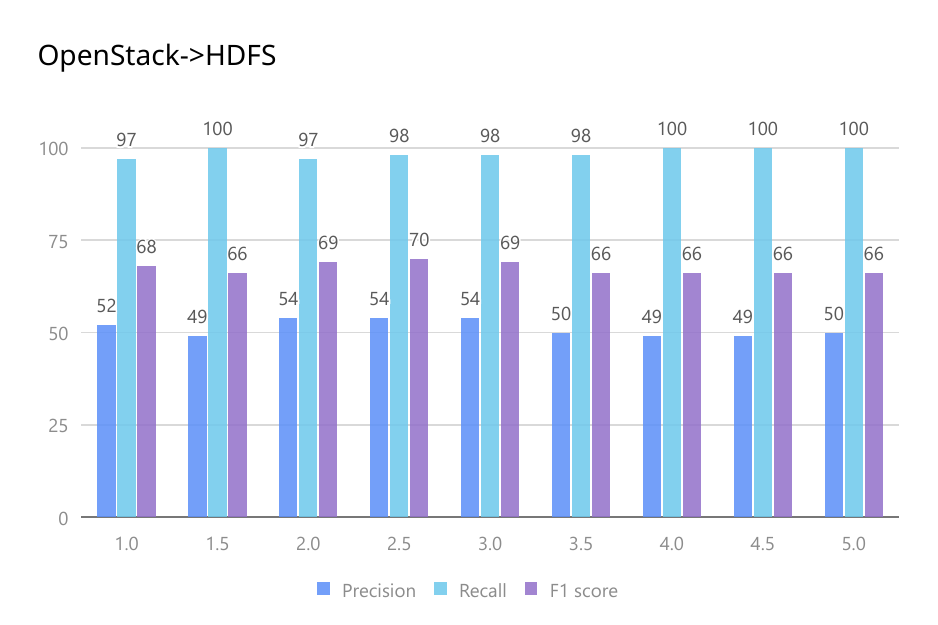}
    \end{minipage}
    \hfill
    \begin{minipage}{0.24\linewidth}
        \centering
        \includegraphics[width=\textwidth]{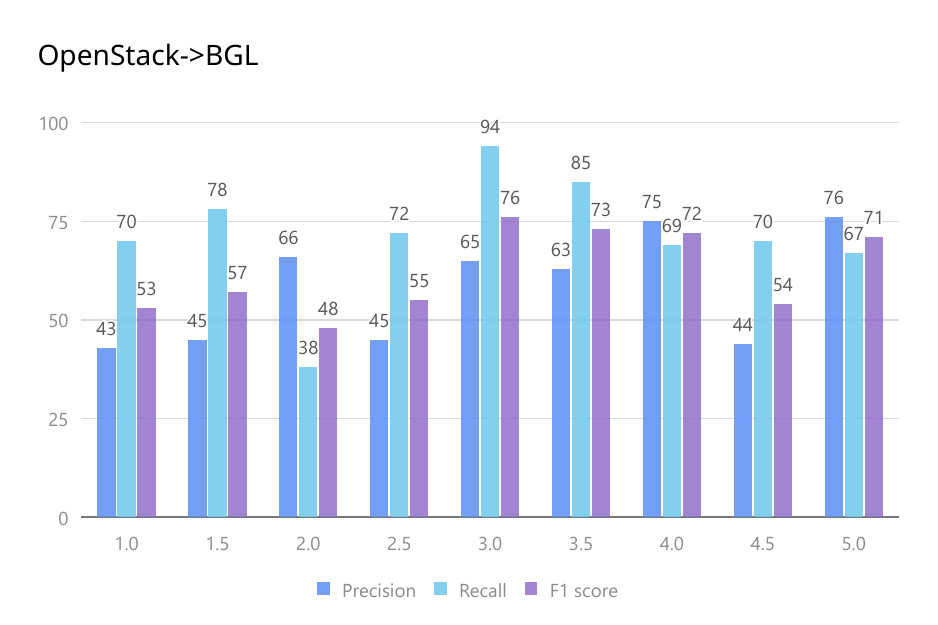}
    \end{minipage}
    \caption{Parameter sensitivity studies on the hyperparameters \( \beta \).} 
    \label{fig:row3}
    
    \vspace{8pt} 

    \begin{minipage}{0.24\linewidth}
        \centering
        \includegraphics[width=\textwidth]{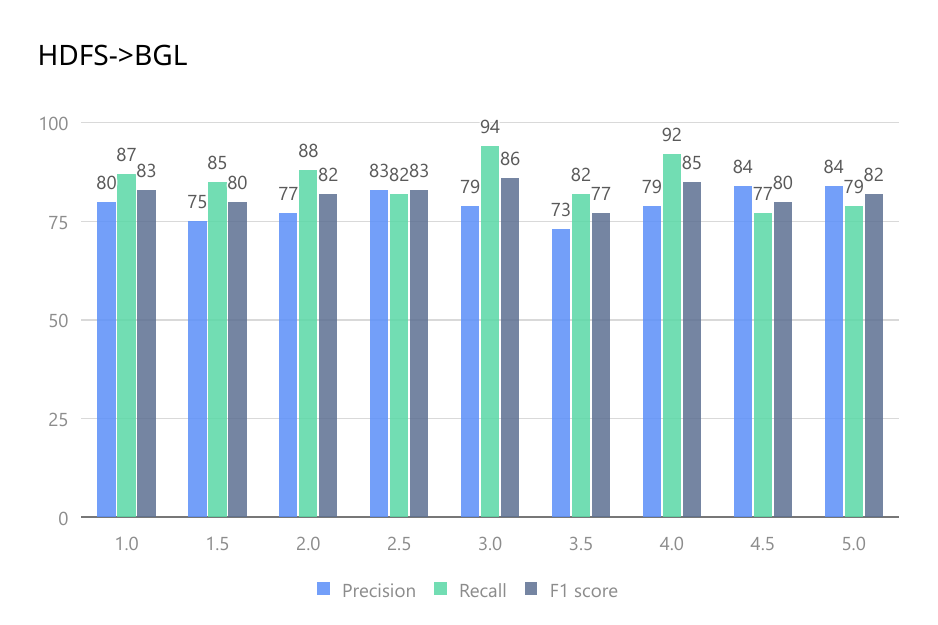}
    \end{minipage}
    \hfill
    \begin{minipage}{0.24\linewidth}
        \centering
        \includegraphics[width=\textwidth]{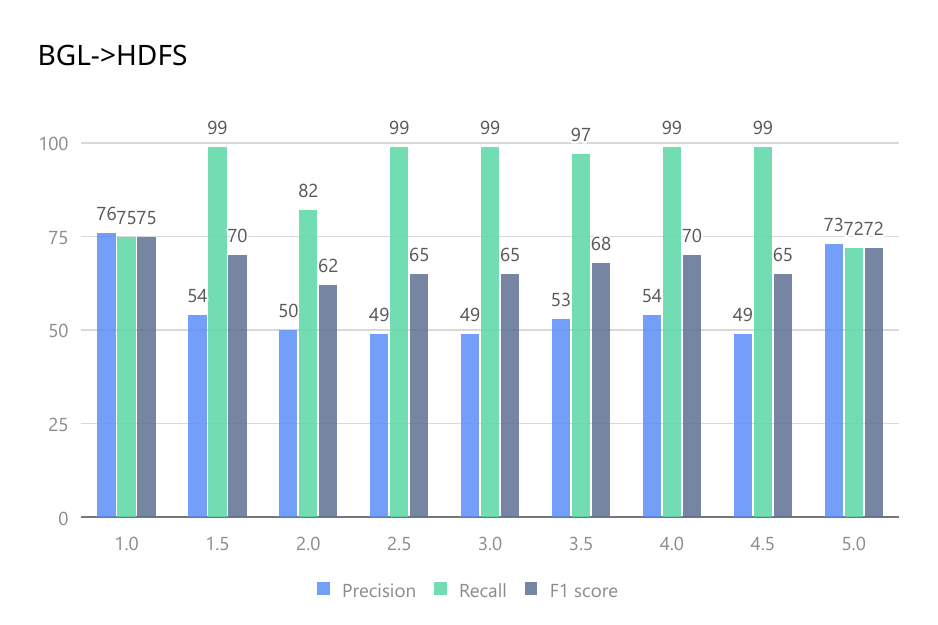}
    \end{minipage}
    \hfill
    \begin{minipage}{0.24\linewidth}
        \centering
        \includegraphics[width=\textwidth]{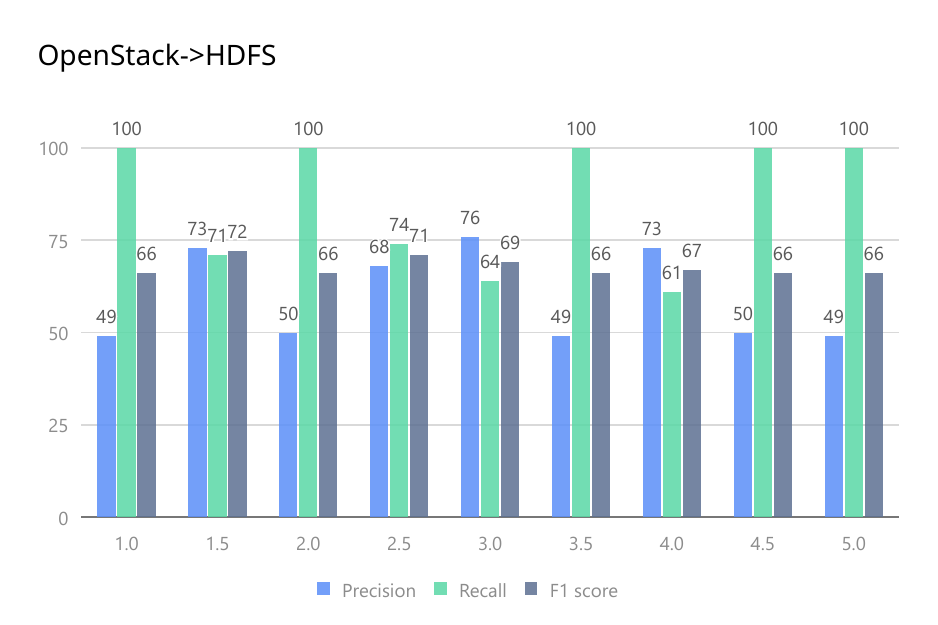}
    \end{minipage}
    \hfill
    \begin{minipage}{0.24\linewidth}
        \centering
        \includegraphics[width=\textwidth]{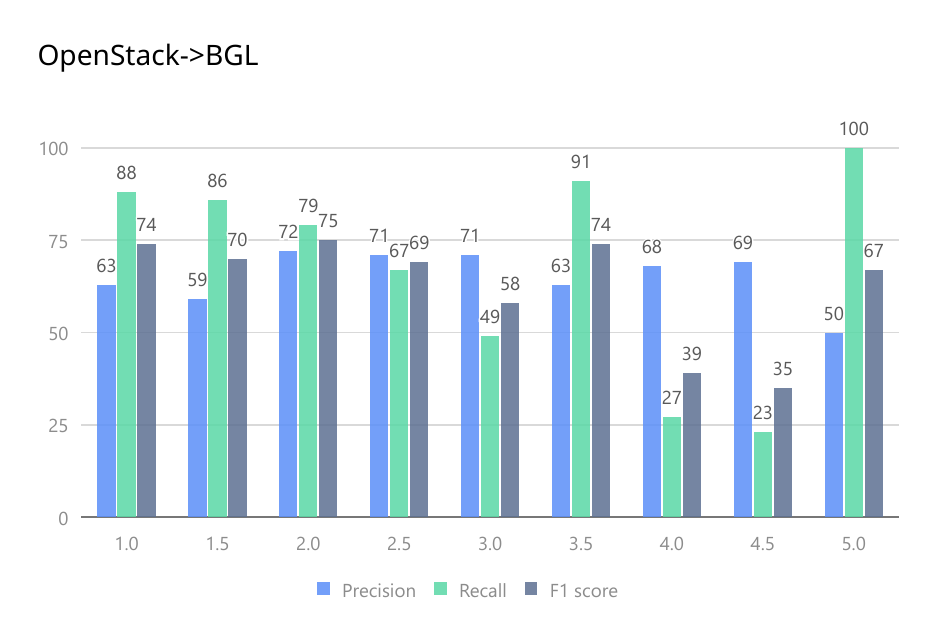}
    \end{minipage}
    \caption{Parameter sensitivity studies on the hyperparameters \( \gamma \).} 
    \label{fig:row4}
\end{figure*}

\subsection{Evaluation on Zero-label Setting}
Zero-label generalization is a highly challenging scenario in cross-system log-based anomaly detection, aiming to evaluate model performance on the target system without utilizing log labels for training. Table \hyperref[tab_all]{2} presents the results of transferring from source system to target system without using target log labels for training. Overall, ZeroLog demonstrates significant advantages over existing cross-system log-based anomaly detection methods in various settings, including direct zero-label scenarios (Block (f)) and transfer learning methods (Block (c)). Furthermore, ZeroLog even outperforms (semi-) supervised methods directly trained on the target log data (Block (a)) in many cases, while substantially surpassing unsupervised methods (Block (e)). Finally, ZeroLog achieves comparable performance to recent methods based on meta-learning (Block (d1)), and it only falls short of (semi-) supervised methods trained with fully annotated (100\%) anomaly labels on the target data (Block (b)). Notably, in contrast to MetaLog (Block (d2) and (d3)), ZeroLog achieves such superior performance without leveraging target log labels for training.  

Specifically, in HDFS to BGL generalization experiment, compared to the state-of-the-art transfer learning method ((c1)-(c3)), ZeroLog improves the F1-score by 18\%, 14\% and 14\%, respectively. This significant improvement indicates that ZeroLog achieves better cross-system performance without utilizing log labels from the target system. Additionally, compared to zero-label direct transfer methods ((f1)–(f4)), ZeroLog achieves at least a 54\% increase in F1-score. When MetaLog uses only the normal labels of the target system, ZeroLog far outperforms its performance ((d2) and (d3)). Remarkably, ZeroLog even outperforms (semi-)supervised methods directly trained on the target data ((a1) and (a2)) with an F1-score improvement of up to 16\%, while surpassing unsupervised methods (e1) by 53\%. These results suggest that source system information enhances the model’s ability to perform better on the target log system. Finally, ZeroLog only slightly lags behind (semi-)supervised methods trained with fully annotated (100\%) anomaly labels ((b1) and (b2)) and recent methods based meta-learning with normal and anomaly labels (d1). 

In the BGL to HDFS generalization experiment, similar results were observed. Compared to existing transfer learning methods ((c1)-(c3)), ZeroLog achieves up to a 33\% improvement in F1-score. Moreover, ZeroLog significantly outperforms zero-label direct transfer methods ((f1)–(f4)) and the unsupervised method (e1). ZeroLog also achieves up to an 18\% improvement in F1-score compared to (semi-)supervised methods ((a1) and (a2)) directly trained on the target data. When MetaLog uses only the normal labels of the target system, ZeroLog achieves an F1-score improvement of at least 84\% ((d2) and (d3)). ZeroLog only falls short of (semi-)supervised methods trained with fully annotated (100\%) anomaly labels ((b1) and (b2)). Notably, compared to HDFS to BGL generalization, ZeroLog performs on par with the method based on meta-learning (d1) in the BGL to HDFS zero-label generalization scenario.  

In the OpenStack to HDFS generalization experiment, ZeroLog achieves up to a 19\% improvement in F1-score compared to existing transfer learning methods ((c1)-(c3)). Additionally, ZeroLog significantly outperforms zero-label direct transfer methods ((f1)–(f4)), the unsupervised method (e1) and MteaLog without using anomaly labels of target system ((d2) and (d3)). ZeroLog even achieves up to a 17\% improvement in F1-score compared to (semi-)supervised methods ((a1) and (a2)) trained directly on the target data. ZeroLog only falls short of (semi-)supervised methods trained with fully annotated (100\%) anomaly labels ((b1) and (b2)). Notably, compared to HDFS to BGL and BGL to HDFS generalization, ZeroLog improves its F1-score by 6\% relative to the method based meta-learning with normal and anomaly labels (d1) in the OpenStack to HDFS zero-label generalization scenario.  

In the OpenStack to BGL generalization experiment, ZeroLog achieves up to a 16\% improvement in F1-score compared to existing transfer learning methods ((c1)-(c3)). Furthermore, ZeroLog significantly outperforms zero-label direct transfer methods ((f1)–(f4)) and the unsupervised method (e1) and MteaLog without using anomaly labels ((d2) and (d3)). ZeroLog performs comparably to (semi-)supervised methods ((a1) and (a2)) directly trained on the target data. ZeroLog only falls short of (semi-)supervised methods trained with fully annotated (100\%) anomaly labels ((b1) and (b2)) and the method based meta-learning with normal and anomaly labels (d1). In contrast, ZeroLog was trained without the use of target log labels, demonstrating its strong zero-label generalization capabilities. 

Overall, existing methods struggle to achieve effective cross-system anomaly detection under zero-label conditions, including the state-of-the-art few-label approach MetaLog. When anomaly labels or all labels are unavailable in the target system, MetaLog based solely on meta-learning fails to extract system-agnostic general feature representations, thus limiting its generalization capability. In contrast, ZeroLog successfully addresses the zero-label cold-start challenge faced by new systems by integrating unsupervised domain adaptation based on adversarial training with meta-learning method.
\subsection{Parameter Sensitivity Analysis}
\label{sec4.3}
We conducted two core sensitivity analysis experiments to investigate the impact of (1) the source and target domain dataset usage ratios during training (as shown in Figures \hyperref[fig:row1]{4} and \hyperref[fig:row2]{5}) and (2) the hyperparameters \(\beta\) and \(\gamma\), which balance adaptability and classification capability (as shown in Figures \hyperref[fig:row3]{6} and \hyperref[fig:row4]{7}), on system performance. The former demonstrates ZeroLog's robustness under various training conditions, while the latter examines how the trade-off between adaptability and classification capability affects performance.

Figure \hyperref[fig:row1]{4} illustrates the influence of the source domain dataset usage ratio on ZeroLog's performance during training. Since the OpenStack dataset is significantly smaller, we focused on experiments using HDFS and BGL as the source domains. The horizontal axis represents the source domain dataset usage ratio (ranging from 10\% to 100\%). In the HDFS to BGL transfer scenario, ZeroLog achieves an F1-score of 80\% even with only 10\% of the HDFS dataset used for training, demonstrating its robustness under limited source domain data. As the usage ratio of the HDFS dataset increases to 90\%, the F1-score improves further to 86\%, achieving exceptional cross-system anomaly detection performance. Figure \hyperref[fig:row2]{5} shows the impact of the target domain's unlabeled dataset usage ratio on ZeroLog's performance. Similarly, in the BGL to HDFS transfer scenario, ZeroLog achieves an F1-score of 76\% with 10\% of the HDFS dataset, which further increases to 81\% with 100\% of the dataset. These results confirm ZeroLog's adaptability and superior performance under varying training data conditions. These findings highlight that the adequate utilization of source and target domain datasets, combined with an effective meta-learning mechanism, enables ZeroLog to better understand log-based anomaly detection tasks and deliver superior cross-system anomaly detection performance.

In ZeroLog network, \(\beta\) and \(\gamma\) are hyperparameters that balance adaptability and classification capability, playing a critical role in moderating the impact of source and target data on network training. Intuitively, a larger \(\beta\) enhances the network's ability to extract shared features between the source and target domains, improving generalization from the source system to the target system. Conversely, a larger \(\gamma\) strengthens the network's capability to classify anomalous logs, enabling it to learn features critical for effective anomaly detection. However, an excessively large \(\beta\) may reduce the network's ability to classify anomalies, adversely affecting anomaly detection performance. Similarly, an overly large \(\gamma\) may compromise generalization performance, hindering effective log-based anomaly detection in the target system. Therefore, appropriate trade-offs in the values of \(\beta\) and \(\gamma\) are necessary. We employed a controlled variable method to demonstrate how variations in the hyperparameters balancing adaptability and classification capability affect performance. Figures \hyperref[fig:row3]{6} and \hyperref[fig:row4]{7} depict the impact of parameter changes on ZeroLog's performance. The horizontal axis represents the parameter range (from 1.0 to 5.0). As shown in the figures, under the zero-label generalization setting from HDFS to BGL, ZeroLog achieves an F1-score of 83\% when \(\beta = 2\) and \(\gamma = 2.5\). Conversely, overly large or small values of \(\beta\) and \(\gamma\) significantly degrade performance, validating the aforementioned intuitive reasoning.

\subsection{Evaluation on Source System}
\label{sec4.4}
In order to verify whether the system‐agnostic general feature representations extracted between the source and target systems inadvertently lose system‐specific characteristics, we conducted single‐domain performance experiments on the source systems. In the zero‐label generalization experiments (HDFS to BGL and BGL to HDFS), we used 80\% of the source system data for training and reserved the remaining 20\% for testing. The results are as follows: on the HDFS source data, ZeroLog achieved a precision of 99.76, a recall of 99.88 and an F1‑score of 99.82; on the BGL source data, it achieved a precision of 99.89, a recall of 99.56 and an F1‑score of 99.73. When evaluated on the corresponding target systems, ZeroLog attained F1‑scores of 82 and 79, respectively (as shown in Figures \hyperref[fig:row1]{4}). Overall, in the course of zero‐label generalization, ZeroLog maintained its outstanding source domain performance while also demonstrating strong generalization capability.
\subsection{Ablation Studies}
\label{sec4.5}
Table \hyperref[tab3]{3} presents the results of our ablation study, which evaluates the contribution of meta‑learning to ZeroLog’s performance. To isolate this effect, we remove the meta‑learning algorithm from ZeroLog and retain only the base model under identical experimental settings. The results demonstrate that, without meta‑learning, the model’s performance deteriorates: in the OpenStack to HDFS experiment, the F1‑score drops to merely 53.70, further underscoring the critical role of meta‑learning in enhancing ZeroLog’s adaptability and anomaly detection effectiveness on target systems.

\begin{table}[ht]
  \centering
  \caption{Experimental results of ZeroLog without meta-learning.}
  \begin{tabular}{lccc}
    \hline
    \textbf{Setting}            & \textbf{Precision} & \textbf{Recall} & \textbf{F1‑score} \\
    \hline
    HDFS to BGL                & 72.04              & 54.02           & 61.74             \\
    BGL to HDFS                & 85.41              & 49.66           & 62.81             \\
    OpenStack to HDFS          & 100.00             & 36.71           & 53.70             \\
    OpenStack to BGL           & 70.39              & 63.64           & 66.84             \\
    \hline
  \end{tabular}
  \label{tab3}
\end{table}

\subsection{Run Time Analysis}
\label{sec4.6}
To validate the time cost and inference efficiency of ZeroLog, we conducted runtime experiments on a single NVIDIA 3090 GPU. The results show that ZeroLog achieves an inference speed of 0.083 ms per input log sequence. In the HDFS to BGL zero‐label generalization experiment, ZeroLog employs a system‐agnostic representation meta‐learning training strategy, which requires approximately 15 minutes for training; inference on the BGL dataset then takes roughly 6 seconds. By contrast, the semi‐supervised method PLELog incurs about 9.5 minutes of training time and 5 seconds of testing time. For the fully supervised baseline LogRobust, since it does not need to generate probabilistic labels, the training time is reduced to 7.5 minutes. Within the class of meta‐learning approaches, MetaLog’s complex gradient‐update procedure results in approximately 13 minutes of training time and 5 seconds of inference. Among transfer‐learning methods, LogTransfer exhibits a training time similar to that of MetaLog (around 13 minutes), whereas LogTAD requires up to 19 minutes for training, primarily due to slower convergence in its unsupervised regime. Overall, compared to MetaLog, ZeroLog incurs a modest increase in both training and inference time, which stems from its complete avoidance of target‐system log labels during training. In the system‑agnostic representation meta‑learning training strategy, we apply adversarial training to the logs of both source and target systems to extract system‑invariant feature representations, which increases computational overhead. However, thanks to ZeroLog’s nested inner and outer loop optimization across meta‑tasks, its overall time cost remains substantially lower than that of LogTAD. Importantly, unlike the methods discussed above, ZeroLog operates in a true cold‑start scenario without any target‑system labels, thereby significantly reducing annotation costs.

\section{Related Works}
\label{sec5}
\subsection{Log-based Anomaly Detection}
Anomaly detection in software systems has garnered widespread attention, and various methods have been proposed to address this challenge~\cite{ DBLP:journals/corr/abs-1803-04967}. Since logs comprehensively describe a wide range of events within software systems, numerous methods based on deep learning have been introduced in recent years for log-based anomaly detection~\cite{10.1145/3133956.3134015, 10248257, 10.1145/3377813.3381371, ijcai2019p658, 9401970, 9240683, 10.1145/3338906.3338931}. Deeplog~\cite{10.1145/3133956.3134015} utilizes LSTM networks to extract normal patterns from the index sequences of log templates, identifying anomalies when new logs deviate from the offline-learned normal patterns. LogAnomaly~\cite{ijcai2019p658}, on the other hand, enhances anomaly detection by leveraging Word2Vec embedding techniques to extract sequence and quantitative features from log templates. LogRobust~\cite{10.1145/3338906.3338931} employs TF-IDF and word vectorization techniques to convert logs into semantic vectors. In this way, updated logs can be transformed into semantic vectors and incorporated into the model’s training and inference process. PLELog~\cite{9401970} introduces a semi-supervised learning model that derives data labels through unsupervised clustering methods, which are then used to build a supervised anomaly detection model. NeuralLog~\cite{9678773} avoids errors and information loss that may occur in traditional log parsing by directly processing raw logs. This method uses deep learning techniques to understand and analyze patterns and anomalous behaviors in log data. KnowLog~\cite{10.1145/3597503.3623304} enhances the understanding of specific terms, abbreviations, and contextual information in logs by incorporating domain knowledge, thereby improving the accuracy and efficiency of log analysis. LogFormer~\cite{10.1609/aaai.v38i1.27764} combines pre-training and adapter fine-tuning strategies to enhance the model’s generalization ability across different log data. LogTransfer~\cite{9251092} and LogTAD~\cite{10.1145/3459637.3482209} propose transfer learning methods that enable cross-system anomaly detection by sharing neural network components between source and target systems. MetaLog~\cite{10.1145/3597503.3639205} presents a cross-system log-based anomaly detection method based on meta-learning that improves the model's generalization ability across different systems through global consistent semantic embedding and meta-learning techniques. However, transfer learning methods exhibit certain limitations when facing significant distribution differences between the source and target domains. Furthermore, MetaLog still requires annotated logs from the target system to achieve satisfactory anomaly detection results, failing to fully address the cold-start problem. Therefore, these methods are not suitable for application in our scenario.

\subsection{Cross-Domain Learning}
In the research on Unsupervised Domain Adaptation (UDA), the goal is to reduce the distribution discrepancy between the source and target domains, utilizing abundant labeled data from the source domain to classify unlabeled data in the target domain. Methods based on adversarial loss have made significant progress by minimizing the distributional differences between domains to improve adaptation performance. Specifically, adversarial training seeks to learn domain-invariant feature representations that are not only discriminative in the source domain but also difficult to distinguish across domains. For example, the Deep Adversarial Neural Network (DANN)~\cite{pmlr-v37-ganin15} introduces adversarial training to construct a feature space that is invariant to domain shifts, aiming to maintain classification capability while being domain-invariant. Adversarial Discriminative Domain Adaptation (ADDA)~\cite{8099799} further improves model accuracy by learning domain-specific encoders to achieve adversarial adaptation. In recent years, meta-learning has gained widespread attention as an effective learning framework~\cite{pmlr-v70-finn17a, 9428530, DBLP:journals/corr/abs-1802-03596}, and has been extensively applied to domain generalization tasks. The goal of domain generalization is to enable neural networks to perform well in both source and target domains. Since the concept of applying meta-learning to domain generalization was first introduced in~\cite{10.5555/3504035.3504462}, this area has received considerable attention, with significant progress made in several downstream tasks, such as long-tail visual recognition~\cite{Jamal_2020_CVPR} and person re-identification~\cite{Zhao_2021_CVPR}, as well as theoretical advancements in both supervised~\cite{Qiao_2020_CVPR} and unsupervised~\cite{pmlr-v181-narayanan22a} domain generalization. These research approaches provide valuable insights for applying UDA and meta-learning mechanisms to address the zero-label cross-system log-based anomaly detection problem.

\section{Conclusion}
\label{sec6}
In this paper, we study a valuable yet underexplored setting: zero-label generalizable cross-system log-based anomaly detection, and propose a novel log-based anomaly detection method, ZeroLog. To achieve zero-label generalization, we design a system-agnostic representation meta-learning method that effectively leverages anomaly classification features from labeled logs in the source domain, as well as domain-invariant features between the source and target domains, to perform cross-system log-based anomaly detection. In the zero-label generalization scenario, ZeroLog achieves an F1-score exceeding 80$\%$, and it performs comparably to the latest cross-system log-based anomaly detection methods that use labeled logs for training. Furthermore, ZeroLog significantly outperforms existing methods in log-based anomaly detection tasks in the target system under the zero-label setup, demonstrating its potential for generalized anomaly detection across diverse software systems. In future research, we will further test the zero-label generalization performance of ZeroLog on more types of log datasets and explore its practical applicability. Going further, our ultimate goal is to address the most challenging zero-shot generalization scenarios.

\section*{Acknowledgment}
This work was supported by National Key Laboratory of Data Space Technology and System.

\bibliographystyle{IEEEtran}
\bibliography{reference}

@String{Computing = "Computing" }

@String{Computer = "{IEEE} Computer" }

@String{Chelsea = "Chelsea" }

@String{Springer = "Springer-Verlag" }

@ARTICLE{10910212,
  author={Zhou, Junwei and Ying, Shaowen and Wang, Shulan and Zhao, Dongdong and Xiang, Jianwen and Liang, Kaitai and Liu, Peng},
  journal={IEEE Transactions on Dependable and Secure Computing}, 
  title={LogDLR: Unsupervised Cross-System Log Anomaly Detection Through Domain-Invariant Latent Representation}, 
  year={2025},
  volume={22},
  number={4},
  pages={4456-4471},
  doi={10.1109/TDSC.2025.3548050}}

@inproceedings{eagerlog,
  title={EagerLog: Active Learning Enhanced Retrieval Augmented Generation for Log-based Anomaly Detection},
  author={Duan, Chiming and Jia, Tong and Yang, Yong and Liu, Guiyang and Liu, Jinbu and Zhang, Huxing and Zhou, Qi and Li, Ying and Huang, Gang},
  booktitle={ICASSP 2025-2025 IEEE International Conference on Acoustics, Speech and Signal Processing (ICASSP)},
  pages={1--5},
  year={2025},
  organization={IEEE}
}

@inproceedings{midlog,
  title={Weakly-supervised Log-based Anomaly Detection with Inexact Labels via Multi-instance Learning},
  author={He, Minghua and Jia, Tong and Duan, Chiming and Cai, Huaqian and Li, Ying and Huang, Gang},
  booktitle={2025 IEEE/ACM 47th International Conference on Software Engineering (ICSE)},
  pages={726--726},
  year={2025},
  organization={IEEE Computer Society}
}

@inproceedings{afalog,
  title={Afalog: A general augmentation framework for log-based anomaly detection with active learning},
  author={Duan, Chiming and Jia, Tong and Cai, Huaqian and Li, Ying and Huang, Gang},
  booktitle={2023 IEEE 34th International Symposium on Software Reliability Engineering (ISSRE)},
  pages={46--56},
  year={2023},
  organization={IEEE}
}

@inproceedings{FreeLog,
author = {Zhao, Xinlong and Jia, Tong and He, Minghua and Wu, Yihan and Li, Ying and Huang, Gang},
title = {From Few-Label to Zero-Label: An Approach for Cross-System Log-Based Anomaly Detection with Meta-Learning},
year = {2025},
isbn = {9798400712760},
publisher = {Association for Computing Machinery},
address = {New York, NY, USA},
url = {https://doi.org/10.1145/3696630.3728519},
doi = {10.1145/3696630.3728519},
booktitle = {Proceedings of the 33rd ACM International Conference on the Foundations of Software Engineering},
pages = {661–665},
numpages = {5},
location = {Clarion Hotel Trondheim, Trondheim, Norway},
series = {FSE Companion '25}
}

@article{hilog,
  title={Hilogx: noise-aware log-based anomaly detection with human feedback},
  author={Jia, Tong and Li, Ying and Yang, Yong and Huang, Gang},
  journal={The VLDB Journal},
  volume={33},
  number={3},
  pages={883--900},
  year={2024},
  publisher={Springer}
}

@inproceedings{aclog,
  title={Aclog: An approach to detecting anomalies from system logs with active learning},
  author={Duan, Chiming and Jia, Tong and Li, Ying and Huang, Gang},
  booktitle={2023 IEEE International Conference on Web Services (ICWS)},
  pages={436--443},
  year={2023},
  organization={IEEE}
}

@inproceedings{10.1145/3133956.3134015,
    author = {Du, Min and Li, Feifei and Zheng, Guineng and Srikumar, Vivek},
    title = {DeepLog: Anomaly Detection and Diagnosis from System Logs through Deep Learning},
    year = {2017},
    isbn = {9781450349468},
    publisher = {Association for Computing Machinery},
    address = {New York, NY, USA},
    url = {https://doi.org/10.1145/3133956.3134015},
    doi = {10.1145/3133956.3134015},
    booktitle = {Proceedings of the 2017 ACM SIGSAC Conference on Computer and Communications Security},
    pages = {1285–1298},
    numpages = {14},
    location = {Dallas, Texas, USA},
    series = {CCS '17}
}

@inproceedings{10.1145/3338906.3338931,
author = {Zhang, Xu and Xu, Yong and Lin, Qingwei and Qiao, Bo and Zhang, Hongyu and Dang, Yingnong and Xie, Chunyu and Yang, Xinsheng and Cheng, Qian and Li, Ze and Chen, Junjie and He, Xiaoting and Yao, Randolph and Lou, Jian-Guang and Chintalapati, Murali and Shen, Furao and Zhang, Dongmei},
title = {Robust log-based anomaly detection on unstable log data},
year = {2019},
isbn = {9781450355728},
publisher = {Association for Computing Machinery},
address = {New York, NY, USA},
url = {https://doi.org/10.1145/3338906.3338931},
doi = {10.1145/3338906.3338931},
booktitle = {Proceedings of the 2019 27th ACM Joint Meeting on European Software Engineering Conference and Symposium on the Foundations of Software Engineering},
pages = {807–817},
numpages = {11},
location = {Tallinn, Estonia},
series = {ESEC/FSE 2019}
}

@inproceedings{10.1145/3377813.3381371,
author = {Kim, Jinhan and Savchenko, Valeriy and Shin, Kihyuck and Sorokin, Konstantin and Jeon, Hyunseok and Pankratenko, Georgiy and Markov, Sergey and Kim, Chul-Joo},
title = {Automatic abnormal log detection by analyzing log history for providing debugging insight},
year = {2020},
isbn = {9781450371230},
publisher = {Association for Computing Machinery},
address = {New York, NY, USA},
url = {https://doi.org/10.1145/3377813.3381371},
doi = {10.1145/3377813.3381371},
booktitle = {Proceedings of the ACM/IEEE 42nd International Conference on Software Engineering: Software Engineering in Practice},
pages = {71–80},
numpages = {10},
location = {Seoul, South Korea},
series = {ICSE-SEIP '20}
}

@inproceedings{ijcai2019p658,
  title     = {LogAnomaly: Unsupervised Detection of Sequential and Quantitative Anomalies in Unstructured Logs},
  author    = {Meng, Weibin and Liu, Ying and Zhu, Yichen and Zhang, Shenglin and Pei, Dan and Liu, Yuqing and Chen, Yihao and Zhang, Ruizhi and Tao, Shimin and Sun, Pei and Zhou, Rong},
  booktitle = {Proceedings of the Twenty-Eighth International Joint Conference on
               Artificial Intelligence, {IJCAI-19}},
  publisher = {International Joint Conferences on Artificial Intelligence Organization},
  pages     = {4739--4745},
  year      = {2019},
  month     = {7},
  doi       = {10.24963/ijcai.2019/658},
  url       = {https://doi.org/10.24963/ijcai.2019/658},
}

@INPROCEEDINGS{9240683,
  author={Yin, Kun and Yan, Meng and Xu, Ling and Xu, Zhou and Li, Zhao and Yang, Dan and Zhang, Xiaohong},
  booktitle={2020 IEEE International Conference on Software Maintenance and Evolution (ICSME)}, 
  title={Improving Log-Based Anomaly Detection with Component-Aware Analysis}, 
  year={2020},
  volume={},
  number={},
  pages={667-671},
  doi={10.1109/ICSME46990.2020.00069}}

@INPROCEEDINGS{9401970,
  author={Yang, Lin and Chen, Junjie and Wang, Zan and Wang, Weijing and Jiang, Jiajun and Dong, Xuyuan and Zhang, Wenbin},
  booktitle={2021 IEEE/ACM 43rd International Conference on Software Engineering (ICSE)}, 
  title={Semi-Supervised Log-Based Anomaly Detection via Probabilistic Label Estimation}, 
  year={2021},
  volume={},
  number={},
  pages={1448-1460},
  doi={10.1109/ICSE43902.2021.00130}}

@INPROCEEDINGS{8854736,
  author={Xia, Wensheng and Li, Ying and Jia, Tong and Wu, Zhonghai},
  booktitle={2019 IEEE 19th International Conference on Software Quality, Reliability and Security (QRS)}, 
  title={BugIdentifier: An Approach to Identifying Bugs via Log Mining for Accelerating Bug Reporting Stage}, 
  year={2019},
  volume={},
  number={},
  pages={167-175},
  doi={10.1109/QRS.2019.00033}}

@inproceedings{10.1145/3534678.3539106,
author = {Jia, Tong and Li, Ying and Yang, Yong and Huang, Gang and Wu, Zhonghai},
title = {Augmenting Log-based Anomaly Detection Models to Reduce False Anomalies with Human Feedback},
year = {2022},
isbn = {9781450393850},
publisher = {Association for Computing Machinery},
address = {New York, NY, USA},
url = {https://doi.org/10.1145/3534678.3539106},
doi = {10.1145/3534678.3539106},
booktitle = {Proceedings of the 28th ACM SIGKDD Conference on Knowledge Discovery and Data Mining},
pages = {3081–3089},
numpages = {9},
location = {Washington DC, USA},
series = {KDD '22}
}

@INPROCEEDINGS{10248257,
  author={Duan, Chiming and Jia, Tong and Li, Ying and Huang, Gang},
  booktitle={2023 IEEE International Conference on Web Services (ICWS)}, 
  title={AcLog: An Approach to Detecting Anomalies from System Logs with Active Learning}, 
  year={2023},
  volume={},
  number={},
  pages={436-443},
  doi={10.1109/ICWS60048.2023.00062}}

@INPROCEEDINGS{9251092,
  author={Chen, Rui and Zhang, Shenglin and Li, Dongwen and Zhang, Yuzhe and Guo, Fangrui and Meng, Weibin and Pei, Dan and Zhang, Yuzhi and Chen, Xu and Liu, Yuqing},
  booktitle={2020 IEEE 31st International Symposium on Software Reliability Engineering (ISSRE)}, 
  title={LogTransfer: Cross-System Log Anomaly Detection for Software Systems with Transfer Learning}, 
  year={2020},
  volume={},
  number={},
  pages={37-47},
  doi={10.1109/ISSRE5003.2020.00013}}

@inproceedings{10.1145/3459637.3482209,
author = {Han, Xiao and Yuan, Shuhan},
title = {Unsupervised Cross-system Log Anomaly Detection via Domain Adaptation},
year = {2021},
isbn = {9781450384469},
publisher = {Association for Computing Machinery},
address = {New York, NY, USA},
url = {https://doi.org/10.1145/3459637.3482209},
doi = {10.1145/3459637.3482209},
booktitle = {Proceedings of the 30th ACM International Conference on Information \& Knowledge Management},
pages = {3068–3072},
numpages = {5},
location = {Virtual Event, Queensland, Australia},
series = {CIKM '21}
}

@inproceedings{10.1145/3597503.3639205,
author = {Zhang, Chenyangguang and Jia, Tong and Shen, Guopeng and Zhu, Pinyan and Li, Ying},
title = {MetaLog: Generalizable Cross-System Anomaly Detection from Logs with Meta-Learning},
year = {2024},
isbn = {9798400702174},
publisher = {Association for Computing Machinery},
address = {New York, NY, USA},
url = {https://doi.org/10.1145/3597503.3639205},
doi = {10.1145/3597503.3639205},
booktitle = {Proceedings of the IEEE/ACM 46th International Conference on Software Engineering},
articleno = {154},
numpages = {12},
location = {Lisbon, Portugal},
series = {ICSE '24}
}

@ARTICLE{7542175,
  author={Ghifary, Muhammad and Balduzzi, David and Kleijn, W. Bastiaan and Zhang, Mengjie},
  journal={IEEE Transactions on Pattern Analysis and Machine Intelligence}, 
  title={Scatter Component Analysis: A Unified Framework for Domain Adaptation and Domain Generalization}, 
  year={2017},
  volume={39},
  number={7},
  pages={1414-1430},
  doi={10.1109/TPAMI.2016.2599532}}

@inproceedings{ijcai2022p496,
  title     = {A Unified Meta-Learning Framework for Dynamic Transfer Learning},
  author    = {Wu, Jun and He, Jingrui},
  booktitle = {Proceedings of the Thirty-First International Joint Conference on
               Artificial Intelligence, {IJCAI-22}},
  publisher = {International Joint Conferences on Artificial Intelligence Organization},
  editor    = {Lud De Raedt},
  pages     = {3573--3579},
  year      = {2022},
  month     = {7},
  note      = {Main Track},
  doi       = {10.24963/ijcai.2022/496},
  url       = {https://doi.org/10.24963/ijcai.2022/496},
}

@inproceedings{NIPS2014_375c7134,
 author = {Yosinski, Jason and Clune, Jeff and Bengio, Yoshua and Lipson, Hod},
 booktitle = {Advances in Neural Information Processing Systems},
 editor = {Z. Ghahramani and M. Welling and C. Cortes and N. Lawrence and K.Q. Weinberger},
 pages = {},
 publisher = {Curran Associates, Inc.},
 title = {How transferable are features in deep neural networks?},
 url = {https://proceedings.neurips.cc/paper_files/paper/2014/file/375c71349b295fbe2dcdca9206f20a06-Paper.pdf},
 volume = {27},
 year = {2014}
}

@article{DBLP:journals/corr/abs-1802-03596,
  author       = {Fengwei Zhou and
                  Bin Wu and
                  Zhenguo Li},
  title        = {Deep Meta-Learning: Learning to Learn in the Concept Space},
  journal      = {CoRR},
  volume       = {abs/1802.03596},
  year         = {2018},
  url          = {http://arxiv.org/abs/1802.03596},
  eprinttype    = {arXiv},
  eprint       = {1802.03596},
  bibsource    = {dblp computer science bibliography, https://dblp.org}
}

@ARTICLE{9428530,
  author={Hospedales, Timothy and Antoniou, Antreas and Micaelli, Paul and Storkey, Amos},
  journal={IEEE Transactions on Pattern Analysis and Machine Intelligence}, 
  title={Meta-Learning in Neural Networks: A Survey}, 
  year={2022},
  volume={44},
  number={9},
  pages={5149-5169},
  doi={10.1109/TPAMI.2021.3079209}}

@inproceedings{NEURIPS2020_cfee3986,
 author = {Goldblum, Micah and Fowl, Liam and Goldstein, Tom},
 booktitle = {Advances in Neural Information Processing Systems},
 editor = {H. Larochelle and M. Ranzato and R. Hadsell and M.F. Balcan and H. Lin},
 pages = {17886--17895},
 publisher = {Curran Associates, Inc.},
 title = {Adversarially Robust Few-Shot Learning: A Meta-Learning Approach},
 url = {https://proceedings.neurips.cc/paper_files/paper/2020/file/cfee398643cbc3dc5eefc89334cacdc1-Paper.pdf},
 volume = {33},
 year = {2020}
}

@inproceedings{gu-etal-2018-meta,
    title = "Meta-Learning for Low-Resource Neural Machine Translation",
    author = "Gu, Jiatao  and
      Wang, Yong  and
      Chen, Yun  and
      Li, Victor O. K.  and
      Cho, Kyunghyun",
    editor = "Riloff, Ellen  and
      Chiang, David  and
      Hockenmaier, Julia  and
      Tsujii, Jun{'}ichi",
    booktitle = "Proceedings of the 2018 Conference on Empirical Methods in Natural Language Processing",
    month = oct # "-" # nov,
    year = "2018",
    address = "Brussels, Belgium",
    publisher = "Association for Computational Linguistics",
    url = "https://aclanthology.org/D18-1398/",
    doi = "10.18653/v1/D18-1398",
    pages = "3622--3631"
}

@INPROCEEDINGS{8029742,
  author={He, Pinjia and Zhu, Jieming and Zheng, Zibin and Lyu, Michael R.},
  booktitle={2017 IEEE International Conference on Web Services (ICWS)}, 
  title={Drain: An Online Log Parsing Approach with Fixed Depth Tree}, 
  year={2017},
  volume={},
  number={},
  pages={33-40},
  doi={10.1109/ICWS.2017.13}}

@inproceedings{10.1145/1629575.1629587,
author = {Xu, Wei and Huang, Ling and Fox, Armando and Patterson, David and Jordan, Michael I.},
title = {Detecting large-scale system problems by mining console logs},
year = {2009},
isbn = {9781605587523},
publisher = {Association for Computing Machinery},
address = {New York, NY, USA},
url = {https://doi.org/10.1145/1629575.1629587},
doi = {10.1145/1629575.1629587},
booktitle = {Proceedings of the ACM SIGOPS 22nd Symposium on Operating Systems Principles},
pages = {117–132},
numpages = {16},
location = {Big Sky, Montana, USA},
series = {SOSP '09}
}

@INPROCEEDINGS{4273008,
  author={Oliner, Adam and Stearley, Jon},
  booktitle={37th Annual IEEE/IFIP International Conference on Dependable Systems and Networks (DSN'07)}, 
  title={What Supercomputers Say: A Study of Five System Logs}, 
  year={2007},
  volume={},
  number={},
  pages={575-584},
  doi={10.1109/DSN.2007.103}}

@INPROCEEDINGS{9678773,
  author={Le, Van-Hoang and Zhang, Hongyu},
  booktitle={2021 36th IEEE/ACM International Conference on Automated Software Engineering (ASE)}, 
  title={Log-based Anomaly Detection Without Log Parsing}, 
  year={2021},
  volume={},
  number={},
  pages={492-504},
  doi={10.1109/ASE51524.2021.9678773}}

@inproceedings{pennington-etal-2014-glove,
    title = "{G}lo{V}e: Global Vectors for Word Representation",
    author = "Pennington, Jeffrey  and
      Socher, Richard  and
      Manning, Christopher",
    editor = "Moschitti, Alessandro  and
      Pang, Bo  and
      Daelemans, Walter",
    booktitle = "Proceedings of the 2014 Conference on Empirical Methods in Natural Language Processing ({EMNLP})",
    month = oct,
    year = "2014",
    address = "Doha, Qatar",
    publisher = "Association for Computational Linguistics",
    url = "https://aclanthology.org/D14-1162/",
    doi = "10.3115/v1/D14-1162",
    pages = "1532--1543"
}

@article{DBLP:journals/corr/abs-1803-04967,
  author       = {Andy Brown and
                  Aaron Tuor and
                  Brian Hutchinson and
                  Nicole Nichols},
  title        = {Recurrent Neural Network Attention Mechanisms for Interpretable System
                  Log Anomaly Detection},
  journal      = {CoRR},
  volume       = {abs/1803.04967},
  year         = {2018},
  url          = {http://arxiv.org/abs/1803.04967},
  eprinttype    = {arXiv},
  eprint       = {1803.04967},
  bibsource    = {dblp computer science bibliography, https://dblp.org}
}

@inproceedings{10.1145/3597503.3623304,
author = {Ma, Lipeng and Yang, Weidong and Xu, Bo and Jiang, Sihang and Fei, Ben and Liang, Jiaqing and Zhou, Mingjie and Xiao, Yanghua},
title = {KnowLog: Knowledge Enhanced Pre-trained Language Model for Log Understanding},
year = {2024},
isbn = {9798400702174},
publisher = {Association for Computing Machinery},
address = {New York, NY, USA},
url = {https://doi.org/10.1145/3597503.3623304},
doi = {10.1145/3597503.3623304},
booktitle = {Proceedings of the IEEE/ACM 46th International Conference on Software Engineering},
articleno = {32},
numpages = {13},
location = {Lisbon, Portugal},
series = {ICSE '24}
}

@inproceedings{10.1609/aaai.v38i1.27764,
author = {Guo, Hongcheng and Yang, Jian and Liu, Jiaheng and Bai, Jiaqi and Wang, Boyang and Li, Zhoujun and Zheng, Tieqiao and Zhang, Bo and Peng, Junran and Tian, Qi},
title = {LogFormer: a pre-train and tuning pipeline for log anomaly detection},
year = {2025},
isbn = {978-1-57735-887-9},
publisher = {AAAI Press},
url = {https://doi.org/10.1609/aaai.v38i1.27764},
doi = {10.1609/aaai.v38i1.27764},
booktitle = {Proceedings of the Thirty-Eighth AAAI Conference on Artificial Intelligence and Thirty-Sixth Conference on Innovative Applications of Artificial Intelligence and Fourteenth Symposium on Educational Advances in Artificial Intelligence},
articleno = {16},
numpages = {9},
series = {AAAI'24/IAAI'24/EAAI'24}
}

@inproceedings{10.5555/3504035.3504462,
author = {Li, Da and Yang, Yongxin and Song, Yi-Zhe and Hospedales, Timothy M.},
title = {Learning to generalize: meta-learning for domain generalization},
year = {2018},
isbn = {978-1-57735-800-8},
publisher = {AAAI Press},
booktitle = {Proceedings of the Thirty-Second AAAI Conference on Artificial Intelligence and Thirtieth Innovative Applications of Artificial Intelligence Conference and Eighth AAAI Symposium on Educational Advances in Artificial Intelligence},
articleno = {427},
numpages = {8},
location = {New Orleans, Louisiana, USA},
series = {AAAI'18/IAAI'18/EAAI'18}
}

@InProceedings{Jamal_2020_CVPR,
author = {Jamal, Muhammad Abdullah and Brown, Matthew and Yang, Ming-Hsuan and Wang, Liqiang and Gong, Boqing},
title = {Rethinking Class-Balanced Methods for Long-Tailed Visual Recognition From a Domain Adaptation Perspective},
booktitle = {Proceedings of the IEEE/CVF Conference on Computer Vision and Pattern Recognition (CVPR)},
month = {June},
year = {2020}
}

@InProceedings{Zhao_2021_CVPR,
    author    = {Zhao, Yuyang and Zhong, Zhun and Yang, Fengxiang and Luo, Zhiming and Lin, Yaojin and Li, Shaozi and Sebe, Nicu},
    title     = {Learning to Generalize Unseen Domains via Memory-based Multi-Source Meta-Learning for Person Re-Identification},
    booktitle = {Proceedings of the IEEE/CVF Conference on Computer Vision and Pattern Recognition (CVPR)},
    month     = {June},
    year      = {2021},
    pages     = {6277-6286}
}

@InProceedings{Qiao_2020_CVPR,
author = {Qiao, Fengchun and Zhao, Long and Peng, Xi},
title = {Learning to Learn Single Domain Generalization},
booktitle = {Proceedings of the IEEE/CVF Conference on Computer Vision and Pattern Recognition (CVPR)},
month = {June},
year = {2020}
}

@InProceedings{pmlr-v181-narayanan22a,
  title = 	 {On Challenges in Unsupervised Domain Generalization},
  author =       {Narayanan, Vaasudev and Deshmukh, Aniket Anand and Dogan, Urun and Balasubramanian, Vineeth N.},
  booktitle = 	 {NeurIPS 2021 Workshop on Pre-registration in Machine Learning},
  pages = 	 {42--58},
  year = 	 {2022},
  editor = 	 {Albanie, Samuel and Henriques, Joao F. and Bertinetto, Luca and Hernandez-Garcia, Alex and Doughty, Hazel and Varol, Gul},
  volume = 	 {181},
  series = 	 {Proceedings of Machine Learning Research},
  month = 	 {13 Dec},
  publisher =    {PMLR},
  url = 	 {https://proceedings.mlr.press/v181/narayanan22a.html}
}

@InProceedings{pmlr-v70-finn17a,
  title = 	 {Model-Agnostic Meta-Learning for Fast Adaptation of Deep Networks},
  author =       {Chelsea Finn and Pieter Abbeel and Sergey Levine},
  booktitle = 	 {Proceedings of the 34th International Conference on Machine Learning},
  pages = 	 {1126--1135},
  year = 	 {2017},
  editor = 	 {Precup, Doina and Teh, Yee Whye},
  volume = 	 {70},
  series = 	 {Proceedings of Machine Learning Research},
  month = 	 {06--11 Aug},
  publisher =    {PMLR},
  url = 	 {https://proceedings.mlr.press/v70/finn17a.html}
}

@InProceedings{pmlr-v37-ganin15,
  title = 	 {Unsupervised Domain Adaptation by Backpropagation},
  author = 	 {Ganin, Yaroslav and Lempitsky, Victor},
  booktitle = 	 {Proceedings of the 32nd International Conference on Machine Learning},
  pages = 	 {1180--1189},
  year = 	 {2015},
  editor = 	 {Bach, Francis and Blei, David},
  volume = 	 {37},
  series = 	 {Proceedings of Machine Learning Research},
  address = 	 {Lille, France},
  month = 	 {07--09 Jul},
  publisher =    {PMLR},
  url = 	 {https://proceedings.mlr.press/v37/ganin15.html}
}

@INPROCEEDINGS{8099799,
  author={Tzeng, Eric and Hoffman, Judy and Saenko, Kate and Darrell, Trevor},
  booktitle={2017 IEEE Conference on Computer Vision and Pattern Recognition (CVPR)}, 
  title={Adversarial Discriminative Domain Adaptation}, 
  year={2017},
  volume={},
  number={},
  pages={2962-2971},
  doi={10.1109/CVPR.2017.316}}

\end{document}